\documentclass[12pt,preprint]{aastex}

\begin{document}

\title{Magnetic Effects at the Edge of the Solar System: MHD Instabilities, the {\it de Laval nozzle} Effect and an Extended Jet}

\author{M. Opher\altaffilmark{1} and P. C. Liewer\altaffilmark{1}}

\affil{Jet Propulsion Laboratory, MS 169-506, 4800 Oak Grove Drive, Pasadena, CA 91109}

\email{merav.opher@jpl.nasa.gov}

\author{M. Velli\altaffilmark{1,2}, L. Bettarini\altaffilmark{2}}

\affil{Departament of Astronomy and Space Science, Univ. of Firenzi, Firenzi, IT 50125}

\author{T. I. Gombosi\altaffilmark{3}, W. Manchester 
\altaffilmark{3}, D. L. DeZeeuw \altaffilmark{3}, G. Toth\altaffilmark{3} and I. Sokolov \altaffilmark{3}}\affil{Space Physics Research Laboratory, Department of Atmospheric, 
Ann Arbor, MI}

\begin{abstract}
To model the interaction between the solar wind and the 
interstellar wind, magnetic fields must be included. Recently Opher et al.  
2003 found that, by including the solar magnetic field in a 3D high 
resolution simulation using the University of Michigan BATS-R-US code, 
a jet-sheet structure forms beyond the solar wind Termination Shock. 
Here we present an even higher resolution 
three-dimensional case where the jet extends for $150AU$ beyond the 
Termination Shock. We discuss the formation of the jet due to a de Laval nozzle effect and it's subsequent large period 
oscillation due to magnetohydrodynamic instabilities. To verify the source of the 
instability, we also perform a 
simplified two dimensional-geometry magnetohydrodynamic
calculation of a plane fluid jet embedded in a neutral sheet with the
profiles taken from our 3D simulation. We find remarkable agreement 
with the full three-dimensional evolution. We compare both simulations and the temporal evolution 
of the jet showing that the sinuous mode is the dominant mode that develops into 
a velocity-shear-instability with a growth rate 
of $5 \times 10^{-9} sec^{-1}=0.027~years^{-1}$. As a result, the outer edge of the 
heliosphere presents remarkable dynamics, such as turbulent 
flows caused by the motion of the jet. Further study, e.g., including 
neutrals and the tilt of the solar rotation from the magnetic axis, 
is required before we can definitively address how this outer 
boundary behaves. Already, however, we can say that the magnetic 
field effects are a major player in this region changing our previous 
notion of how the solar system ends.
\end{abstract}


\section{Introduction}
One of the most exciting phenomena in astrophysical media are jets. 
Usually, they are collimated structures extending for several parsecs 
near black holes and active galaxies \citep{jets}. Recently, we 
showed \citep{opher} that due the compression of the solar 
magnetic field, a jet-sheet structure forms
in the region beyond the Termination Shock, 
where the solar wind becomes subsonic. 
This structure is narrow and collimated in the 
meridional plane (jet), but extends in the equatorial plane as a disk 
for approximately $60{}^{o}$ around the upstream direction. This result was 
obtained using the state-of-the art three 
dimensional magnetohydrodynamic code (BATS-R-US) with an unprecedented resolution to 
investigate the region where the solar wind meets the 
interstellar medium. It was motivated by the approach of the Voyager 
spacecraft to this region. Voyager is one of the most successful 
missions of NASA, traveling for more than 25 years with two 
spacecraft, respectively with a velocity of $3.5AU/year$ (35${}^{o}$ 
above the ecliptic plane) and $3.1AU/year$ (48${}^{o}$ below the 
ecliptic plane). 

In this section we review the current state of modeling the 
interaction between the solar and the interstellar wind and discuss 
why magnetic effects are crucial, especially in the region between 
the Termination Shock and the Heliopause, the Heliosheath.

Because the solar system travels in the interstellar 
medium with a velocity of approximately $25km/s$ \citep{frisch}, 
the solar wind collides with a supersonic
interstellar wind. The basic structures that are formed by this 
interaction are: the Termination Shock (TS), the Heliopause (HP), and, 
possibly a Bow 
Shock (BS). The Termination Shock is the boundary where the supersonic 
solar wind becomes subsonic slowing down approaching the Heliopause. 
The Heliopause is the boundary separating the subsonic solar outflow 
downstream of the Termination Shock and the interstellar plasma 
flowing around it. The region inside the Heliopause is called the Heliosphere. 
If the flow of the interstellar wind is 
also superalfvenic, there is a BS further out, where the
interstellar flow becomes subsonic to avoid the heliospheric obstacle.
In previous hydrodynamic models, the region beyond the Termination Shock has  
constant plasma pressure and temperature and the heliospheric boundary was a smooth, 
rounded surface. This is not true if the solar magnetic field is included: the plasma 
pressure, temperature and density downstream the shock are not uniform and constant, and 
the heliospheric boundary is highly distorted from the rounded appearence of the hydrodynamic models.

The region between the Termination Shock and the Heliopause, the
Heliosheath, is one of the most mysterious and unknown regions. 
The total plasma beta (the ratio between the sum of the thermal and ram pressure and 
the magnetic pressure) decreases as one passes through the shock; it is clear that 
the magnetic field will play a major role in that region (Before the TS, 
$\beta_{total} \approx 65$. After the TS $\beta_{total} \approx 8$. At the magnetic 
ridges $\beta_{total} \leq 1 \approx 0.1$). The solar magnetic field reverses polarity at the 
heliospheric current sheet (HCS). If we neglect the tilt between the solar 
magnetic and rotation axis, the HCS remains in the equatorial plane. 
One of the major questions is how the HCS behaves beyond the 
Termination Shock, or in another words, how the current system in the 
Heliosphere closes.

The interaction between the solar and interstellar wind 
is inherently a three dimensional problem because of the magnetic fields. 
Ideally, modeling this complex region properly would require inclusion of 
several ingredients: 
a) the ionized component from the solar 
and interstellar wind; b) the neutral atoms (coupled to the 
plasma by charge-exchange collisions); c) the magnetic field (of both 
the solar wind and the interstellar wind); d) the pickup ions; and e) 
cosmic rays (both galactic and anomalous components). 

For the interstellar wind, we know with certainty neither 
the magnitude nor the direction of the magnetic field. In the case of the 
neutral component, the associated particle large mean free path 
dictates a kinetic treatment. In addition, 
the diverse length scales present in the problem must be treated 
consistently (i.e. the numerical resolution of a simulation code 
must be capable of widely differing resolutions according to the 
different regions). The scales that we are 
interested in are on the order of tenths of an AU (for MHD instabilities 
and the heliospheric current sheet, for example) although the 
dimensions of the Heliosphere are on the order of a thousand AU. 
Therefore, another aspect that is important when modeling this 
interaction is the numerical resolution.

There have been several numerical approaches to tackle this 
complicated interaction. Much work has focused on the careful 
treatment of the neutral component.
\citet{baranov93,baranov95} (see also \citet{baranov98}) developed a 
kinetic description for the neutrals using a Monte Carlo approach. 
These studies, however, neglected magnetic fields. More recently, 
\citet{zaitsev} and \citet{izmodenov} used the same description and included the 
interstellar magnetic field in a two dimensional model. 
Another alternative approach, still in two dimensions and neglecting 
the magnetic field, 
was developed by \citet{pauls} (see also \citep{zank,zankpauls}) who 
treated the neutrals as an additional fluid species. 
More recently, \citet{muller} used a 
particle-in-mesh approach as a kinetic approximation for the 
neutrals. \citet{liewer} and \citet{liewer2} used a gasdynamic approach treating the neutrals as a 
single fluid, also in a two dimensional study. The works by 
\citet{ratkiweicz98} (see also \citep{ratkiweicz00,ratkiweicz03}) and 
\citep{ratkiweiczjaffel} were three-dimensional studies assuming 
that the neutrals have constant density, velocity and temperature. They 
also neglected the solar magnetic field and studied the effect of 
the interstellar magnetic field for several magnitudes and directions.

The works by \citet{pogorolev}, \citet{mcnutt}, 
\citet{washimi,washimiII} and \citet{linde} were the only ones that 
included both the solar and interstellar magnetic fields in a self consistent way in 
three-dimensional treatments. The neutrals were included as a single 
fluid. The major drawback of these studies are: the inclusion of 
both of the interstellar and solar magnetic fields and the neutrals making 
it difficult to differentiate the effect of each component; and the limited spatial resolution.

In our recent study \citep{opher} we performed a three dimensional 
study including the solar magnetic field. In order to study how the HCS 
behaves beyond the TS, we did not include the 
interstellar magnetic field to eliminate the effect of 
reconnection between the solar and interstellar fields and to isolate the effect of the 
solar magnetic field. We used an adaptive mesh refinement allowing us 
to get to spatial resolutions previously not obtained of the order of 
$1.5~AU$ at the HCS. We showed that a {\it jet-sheet} structure  forms 
at the edge of the solar system. This jet-sheet oscillates up and 
down due to a velocity shear instability. In that study, 
we assume as a first approximation that the solar 
magnetic and the rotation axis are aligned. We also only treat the 
ionized component and neglect the effect of neutrals. Both effects, 
i.e., the inclusion of neutrals and the tilt of the dipole, need to be 
included to assess the overall structure and dynamics of the edge of our 
solar system. Spatial resolution was a key factor for resolving the jet-sheet structure 
at the edge of the solar system. We showed that resolutions of 
the order of 3 AU are insufficient; the jet at the current sheet is broadened and 
the current sheet remains in the equatorial plane as in the previous 
studies of \citet{linde,washimi}.  

Figure 1 shows the pressure contours of a meridional cut of our 
simulation with and without the inclusion of the solar magnetic field. 
The coordinate system is such that the interstellar wind is flowing from the negative 
$x$-direction and the solar rotation axis is in the positive $z$-direction. 
The run presented in 
both cases is the {\it coarse} grid case (see \citet{opher} and Appendix) with 
resolution of $\sim 3~AU$ at the HCS. It is clear that in the 
Heliosheath the pressure is not uniform anymore, being enhanced in the 
region of the current sheet. Also, the HP is more pointed and the 
distance between the Heliopause and the Bow Shock increases (the HP location moves from $240~AU$ to $280~AU$ 
and the BS location moves from $358~AU$ to $440~AU$), 
due to the presence of the jet.

In the present study we revisit the work of \citet{opher} in order to 
study the formation of the jet and the magnetohydrodynamic (MHD) instabilities driven 
by the jet-sheet. 
We present a run where we extended the resolution to $0.75~AU$ at the 
HCS. (In the previous 
study we used a resolution of $1.5~AU$ at the current sheet, 3-4 
times more refined than previous studies.) 
With such high resolution and extending the refined region, we 
were able to resolve the jet extending to $150~AU$ beyond the 
Termination Shock. We discuss on the effect of the de Laval nozzle on 
the formation of the jet and the width of the jet. 

The outline of the 
paper is the following: In the first section we describe briefly the 
model (for more details, see \citet{opher}). In the second section, 
we present our major results for the run used and discuss the 
effect of the {\it de Laval nozzle} on formation of the jet. In the third section we discuss the width of the jet. 
The fourth section is dedicated to the discussion of the role of magnetohydrodynamic 
instabilities and comparison of the 3D and 2D MHD code. The fifth section is dedicated to the 
conclusions and discussions. Finally, in the appendix we discuss the dependence of the width of the jet on 
resolution, showing that at the high resolution case, we obtained grid converged numerical solution.

\section{Description of the Model}
The code that we are using, BATS-R-US (Block Adaptive Tree Solar wind Roe type 
Upwind Scheme), was developed at the 
University of Michigan\citep{powell}.The BATS-R-US code has been designed to capitalize on 
advances on massive parallel computers; the evolution of adaptive mesh refinement (AMR) 
techniques and advances in basic numerical methods particularly for 
hyperbolic conservations laws. The BATS-R-US MHD algorithm uses an upwind methods, approximate 
Riemann solvers, and limited solution reconstruction (see \citet{powell}). It is a three dimensional magnetohydrodynamic 
code and the governing equations are the ideal magnetohydrodynamic equations. 
The resulting finite-volume scheme solves for the hydrodynamic and electromagnetic 
effects in a tightly coupled matter. It 
uses block-based solution for the adaptation on a cartesian mesh and 
is able to have more that 20 levels of refinement. It is a parallel 
multi-processor code with capability of running in different 
platforms, such as Beowulf (type PC clusters) and SGI origin 2000 machines.

In this study, as in \citet{opher}, we only treat the ionized 
components neglecting the neutrals. In order to eliminate the effect of reconnection at the 
heliopause, we did not include an interstellar magnetic field. The inner boundary was chosen at 
$30~AU$ and the outer boundary at $3450 \times 4500 \times 4500~AU$. 
The physical boundary conditions are: For the solar wind (at $30~AU$): 
$n_{plasma}=7.8\times 10^{-3}~cm^{-3}$, $T=1.6\times 10^{3}~K$, 
$v=450~km/s$. The solar magnetic field at the inner boundary was 
taken as the Parker spiral magnetic field,
\begin{equation}
\vec{B}=sign(cos \Theta) \left [ B_{0} 
{\left( \frac{R_{0}}{r} \right ) }^{2} \vec{e_{r}} -
B_{0} \left( \frac{{R_{0}}^{2}}{r} \right ) 
\frac{\Omega_{\odot}sin \Theta}{u_{sw}} \vec{e_{\phi}} \right ] ~,
\end{equation}
where $R_{0}$ is taken as the inner boundary $30~AU$, $u_{sw}$ is the 
solar wind speed, $450~km/s$, $\Theta$ is 
the polar angle of the field line, and $\Omega_{\odot}$ is the equatorial 
angular velocity of the Sun. At the equatorial plane,
$B=B_{0}=2\mu G$. We chose the field polarity ($sign(cos \Theta)$) 
to match the 1996 solar minimum. For the interstellar wind: 
$n_{plasma}=0.07 cm^{-3}$, $T=10^{4}~K$, $v=25~km/s$.

\subsection{Jet formation and properties: magnetic ridges and the de 
Laval nozzle}
In the present study we use higher spatial resolutions than the 
previously reported run \citep{opher} obtaining a resolution up to $0.75~AU$ at 
the current sheet. In the 
Appendix we comment on the effect of spatial resolution on the width of the jet. 
In this section, we briefly summarize the major 
results of this run indicating the major features and differences from 
the previous case (with resolution of $1.5~AU$ at the HCS). Figure 2a 
shows the contours of the magnetic field at $t=1.3 \times 10^{9}~sec$ 
(scale ranging from $0-0.3~nT$). The coordinate system is the same as in Figure 1. Beyond the 
the Termination Shock (TS), the azimuthal magnetic field is 
compressed forming {\it magnetic ridges} (strong field, see region in red). 
The spatial resolution at the  {\it magnetic ridges} is $3~AU$, double of the 
resolution of previous works \citep{opher} and the magnetic field 
intensity ranges from $0-0.4~nT$ (an increase from $0.24~nT$ from the 
previous study). Figure 2b shows the contours of the velocity $U_{x}$(in the meridional plane) 
between the TS and the HP. In both, the black lines, are the flow 
streamlines.

The velocity of the jet is $\approx 200~km/s$ extending for $150~AU$ 
beyond the TS, to $x=-320~AU$, in the direction of the BS (located 
at $x=-380~AU$). Due to the higher resolution the turbulent vortices 
are more pronounced (with sizes of $\sim 20~AU$) 
and the structure is more complex than in \citet{opher}. 
The region where the turbulent vortices form has a resolution of 
$1.5-3.0~AU$, while the one in \citet{opher} has a 
resolution of $6.0~AU$ in the same region (see Appendix for details).  
The jet is unstable as in \cite{opher}. We discuss details of the instability in Section 4. 

The jet-sheet structure forms in the region of minimum of magnetic 
pressure. Figure 2c shows that the flow is in pressure 
balance after the shock. It portrays the vertical line cut 
downstream of the TS (at $x=161~AU$), at $y=0$ for thermal (dashed line), magnetic 
(dash-dot-dash line) and total (solid line) pressure. 
Downstream of the shock, where the flow decelerates, conservation of 
magnetic flux outside of the equatorial plane causes the field to
increase its magnitude further, while in the 
current sheet there is no such effect. As a result, the increased magnetic 
field above and below the ecliptic planes effectively pinches the sheet 
just beyond the Termination Shock, causing the stream lines in the 
subsonic regions to converge slightly.
As pointed out by \citet{washimiII}, the enhanced magnetic ridges also
create an obstacle in the post-shock flow,
decelerating and deviating it to the flanks before reaching the Heliopause. 
This can be seen in Figure 2d where a line cut is taken, $60^{\circ}$ 
above the ecliptic plane. The ram pressure (the green curve) reaches a 
minimum at the maximum intensity of the magnetic pressure, the {\it 
magnetic ridges} (red curve). (See also streamlines in Figure 2a).

Figure 3 shows the ratio of velocity of the jet $u$ with respect to 
the sound speed $a$ at $t=1.3 \times 10^{9} sec$, when the current 
sheet is still in the equatorial plane (green curve). For comparison, 
two other cases are presented: the refined (red) and coarse (blue) resolution 
cases. For the coarse case, the de Laval nozzle effect is negligble due to a broader jet. 
As we increase the resolution (red curve), the jet extends 
farther away, in the direction of the Bow Shock. In the refined case 
(as in \citet{opher}) the jet loses it's power 
at $x=-230~AU$ (indicated by the green arrow) while the change in 
resolution (from $1.5$ to $3.0~AU$) occurs at $=-320~AU$. In the extended 
case the jet loses it's power at $x=-320~AU$ (while the BS is at 
$x=-380~AU$, indicated by the red arrow). The change in resolution 
($0.75$ to $1.5~AU$) occurs at $-350~AU$ and, from $1.5$ to $3.0~AU$ at 
$-365~AU$. This indicates that resolution plays a major role and the jet loses 
it's power due to numerical dissipation.

The converging flow lines near the equatorial plane
create a de Laval nozzle: a subsonic flow must accelerate where the streamlines converge and decelerate where flow lines 
diverge, while the opposite is true for a supersonic flow. 
Figure 4 shows a schematic 
drawing of a nozzle along with the streamlines where the jet form. 
The red arrow indicate the region where acceleration occurs. 
For a steady gas flow the Euler's equation can be written as 
\citep{landau} 

\begin{equation}
{\rho v} \frac{{\partial} v}{{\partial} r} = - \frac{{\partial} 
p}{{\partial} r}+F ~,
\end{equation}
where $F$ is the force. Consider a flow of gas through a nozzle with 
a variable cross section $A(r)$ (as pictured in Figure 4a)
Considering the pressure $p=\rho R T$, where $R$ is the gas constant and the mass flux constant 
${\rho} v A(r) =constant$, the equation above can be written as
\begin{equation}
\frac{v^{2} -{c_{s}}^{2}}{v} \frac{dv}{dr} = \frac{T}{{\rho}} 
\frac{1}{A} \frac{dA}{dr} ~.
\end{equation}

For a narrowing nozzle, $dA/dr <0$, and the flow velocity increases as long as 
it is subsonic $v^{2} < {c_{S}}^{2}$, reaching a maximum where the cross section is minimum. 
If the nozzle then widens again, the flow decelerates thereafter.  
However, if the flow accelerates sufficiently
to become supersonic at the neck, then it will continue to accelerate
where the nozzel widens ($dA/dr >0$). 
A nozzle that first narrows and then widens again is 
called the {\it de Laval nozzle}. This is 
analogous to what happens in Parker's model of the solar wind 
at the critical point, where the effective nozzle, created by the
combined effects of mass flux conservation and gravity, has a minimum 
cross section. In the jet formed here, the maximum speed just barely exceeds 
the sound speed, and decelerates thereafter, much as a breeze in solar wind 
theory.

At the TS the flow velocity decreases to Mach number $0.55$. Due to the acceleration of the 
flow past the TS, the Mach number increases to $1.1$. The velocity of the jet is 
$\sim 150~km/s$ and remains  almost constant for the extension of the 
jet (see Figure 3, where the velocity of the jet decreases only at 
$x=-320~AU$ due to the change in resolution). Fig. 4 shows the location of the 
{\it de Laval} nozzle effect.

\section{Width of the Jet}
What determines the scaling of the width of the jet? Does it depend on the 
grid resolution at the current sheet?
Observations from Pioneer and Voyager indicate that the width of the 
HCS, at $1~AU$ is approximately $10,000~km=6.7\times 10^{-5}~AU$ 
(while a surrounding plasma sheet is thicker by a factor of $~30$ 
\citep{smith,daniel}) In terms of proton gyroradius the current sheet 
thickness is about $200~R_{L}$ \citep{daniel}. For all cases, this is smaller than 
our grid resolution.

How does the width of the current sheet scales with distance? 
Neglecting dynamic effects (such as turbulence) and kinetic effects such as drift instabilities, 
and assuming pressure balance, 
the width is determined by a balance between
the thermal (inside the sheet) and the magnetic pressure outside. 
Combining the adiabatic energy and 
the energy equation,

\begin{equation}
( \vec{u} \cdot \vec{\nabla} ) p + \gamma p ( \vec{\nabla} \cdot 
\vec{u} ) = 0~.
\end{equation}

This equation can be written as 
\begin{equation}
u_{0}\frac{dp}{dr}+\gamma p \vec{\nabla} \cdot \vec{u} =0~,
\end{equation}
where $u_{0}$ is the radial solar wind velocity, and {\it r} the radial 
distance from the Sun. 
Let the width of the current sheet be {\it a}, from Eq. (5),
\begin{equation}
\frac{1}{p}\frac{dp}{dr}=\frac{\gamma}{ar}\frac{d(ar)}{dr}
\end{equation}
that gives the dependence for the thermal pressure with distance as
\begin{equation}
p = {(ar)}^{-\gamma}~,
\end{equation}
where for an adiabatic gas $\gamma=5/3$. For a spherical expansion $a 
\propto r$ and the thermal pressure falls as $r^{-10/3}$. However, 
since the azimuthal magnetic field falls with distance as $r^{-1}$, 
the magnetic pressure falls with $r^{-2}$. 
Therefore, from the pressure equilibrium that must exist between the 
thermal pressure at the current sheet and the magnetic pressure 
outside the sheet, the dependence of the width with distance must be,
\begin{equation}
a(r) \propto r^{1/5}~.
\end{equation}
Using this scaling, for the obserevd width of $6.7\times 10^{-5}~AU$ at $1~AU$, 
the current sheet width at $150~AU$ is $2 \times 10^{-4}~AU$ which is 
much smaller than the grid resolution of our most resolved case. 

If the width of the HCS is determined by numerical diffusion, we can estimate 
the width as follows:
For a steady state solution, the induction equation gives the width of the 
current sheet to be
\begin{equation}
\nabla \cdot (\vec{u} \vec{B}-\vec{B}\vec{u}) = \eta _{m}\nabla^{2}{B}
\end{equation}
or approximately 
\begin{equation}
\eta_{m}\frac{B_{\phi}}{a^{2}} \approx u_{z}\frac{B_{\phi}}{a}~,
\end{equation}
giving the approximately the width as 
\begin{equation}
a \approx \frac{\eta_{m}}{u_z}~,
\end{equation} 
i.e., the width of the HCS, {\it a}, in steady state, is determined by numerical diffusion 
$\eta_{m}$ and the converging flows to the current sheet. The numerical diffusion $\eta_{m}$ 
will be proportional to the grid spatial resolution. If numerical diffusion determines width, the 
width decreases as resolution increases. However, this is not observed.

Figure 5a shows the dependence of the jet width on the grid resolution in 
the region of the current sheet ($-10~AU < z < 10~AU$). Three cases 
are included: coarse, refined (same as Opher et al. 2003) and the 
super-refined-extended (the one presented at section 2.1). For the coarse grid, the resolution at the 
current sheet is $6~AU$, for the refined grid, $1.5~AU$ and for the super-refined-extended grid, $0.75~AU$.
The blue and the red curves are, respectively, the half width of the velocity and the magnetic field, 
measured after the TS, at $x=-210~AU$. The half widths of both the 
velocity and the magnetic field are measured half way of the maximum. 
For the three cases, the velocity (jet) profile follows closely the magnetic field (the current sheet)
profile, being always slightly broader. As we increase the resolution at the flanks of the jet, 
the two profiles approaches each other, almost overlapping. (For more details see Appendix). 
As we increase the resolution, the width of the velocity (jet) approaches the width of the 
magnetic field (current sheet). As we increase the resolution, the magnetic field (current sheet) width 
converges to a value of $4~AU$.

Figure 5b presents the the velocity profiles for the three cases: the coarse, the refined and the 
super-refined-extended cases. The red, green and blue curves, are respectively, the super-refined-extended, 
refined and coarse cases. The jet is much more resolved at the flanks as well as at the region of the current sheet 
in the super-refined-extended and the refined cases, compared to the coarse case. 
The profiles for the velocity for the super-refined-extended and the refined case are almost identical. 
We can see that the width of the jet, in the super-refined-extended case, is independent of the grid resolution 
and is determined by the physical conditions, rather than 
by numerical resolution or numerical diffusion.

\section{Instabilities - Comparison of 3D with 2D MHD code}
There have been several studies examining magnetized flows in which 
both sheared flows and sheared magnetic fields are present \citep{einaudi, 
keppens,einaudiII}. A particular configuration studied, motivated by 
the study of the heliospheric current sheet, was of a fluid jet 
embedded in a neutral sheet. Studies like \cite{dahlburg,einaudiII} 
analyzed the behavior of a plane magnetized jet. Both study were 
two-dimensional cases (the justification came from laminar nature of 
the initial system and of the fact that the fastest growing linearly 
unstable modes are 2D). The configuration studied was a fluid jet in 
the $z$ direction and the spatial coordinate along which the mean 
flow varies as $y$: ${\hat W_{0}}(y)=sech (y) \hat{e}_{z}$ embedded in 
a neutral sheet ${\hat B_{0}}=tanh (\delta y) {\hat e}_{z}$, where 
$W_{0}$ and $B_{0}$ are the basic velocity field and magnetic field and 
$\delta$ is the magnetic shear width. 
These studies focus on the response of the wake-neutral or 
jet-neutral sheet to infinitesimal perturbations. The two unstable 
modes that exist in the fluid limit (no magnetic field) are the 
varicose (sausage-like) 
and sinuous (kink-like) modes. These modes differ in the symmetry of
the perturbed velocity component in the direction perpendicular to the 
flow. For the varicose mode, the perturbed velocity component is 
antisymmetric in $y$ and, for the sinuous mode it is symmetric.
In a magnetically dominated limit, only a tearing mode exists. This mode has the same symmetry as the 
varicose fluid mode. In the flow-dominated regime the sinuous mode has the fastest linear growth rate 
(faster than the varicose mode). Both modes are Kelvin-Helmholtz modes of a jet. 
The sinuous and varicose modes become magnetohydrodynamic modes for 
finite values of the Alfven number $A$, the ratio of the 
characteristic Alfven speed to the characteristic flow speed.

In the 3D case presented in Section 2.1, 
$v_{A} \ll 100km/s$ and $v_{flow} \sim 200~km/s$ so the Alfven 
number, $A \ll 1$. At the location of the jet, there is no magnetic field 
and we expect the predominant mode to be 
the sinuous mode.
The jet in the 3D case presented in Section 2.1 is unstable and 
oscillates up and down as in the case presented in 
\citet{opher}, although due to the higher resolution, much more 
structure can be seen (see attached material for a movie of Figure 2a). Figure 
6a presents a measure of the displacement of the current sheet 
as a function of time. The red curve is the displacement in the 
middle of the jet, at $x=-225~AU$, and the green curve the 
displacement at the nose, at $x=-290~AU$. Indexes (I-III) indicate the 
different regions of time scales of oscillation. The displacement at 
the two locations of the jet are anti-correlated, implying that the 
wavelength of the oscillation $\lambda$ is $\leq 70~AU$. Examining 
Fig. 6b it is clear that the oscillation is a sinuous mode and that the 
wavelength is $\lambda \sim 23~AU$. There are two distinct time scales. 
At the beginning of the displacement, a time scale of oscillation of $t_{1}=2\times 
10^{8} sec$ (region I) and then $t_{2}=2\times 10^{9} sec$ (region 
II). Region III marks the exponential growth of the 
velocity-shear-Kelvin-Helmholtz instability. The growth rate, measured for curve 
III is $\Gamma=5.4\times 10^{-9} sec^{-1}$. (As a comparison, the 
growth rate measured for the refined case in \citet{opher} was $1.6 
\times 10^{-9} sec^{-1}$.)

The Kelvin Helmholtz (KH) growth rate, for an infinitely thin layer is \citep{chandra}
$\Gamma =0.5 \times \mid k \cdot U_{0} 
\mid [ 1-{(2c_{A}{\hat k}\cdot {\hat{B}}_{0})}^{2}/{({\hat{k}}\cdot U_{0})}^{2}] ^{1/2}$, where $U_{0}$ is the 
velocity jump across the shear layer, $c_{A}$ is the Alfv\'en speed, ${\hat k}$ is the direction of the 
wavevector (here ${\hat x}$) and ${\hat{B}}_{0}$ is the direction of the magnetic field. Maximum growth occurs for 
$k \approx 1/a$, where $a$ is the distance between the
maximum velocity and zero. For the 3D run performed here, at $x=-190~AU$, 
$U_{0}\approx 150~km/s$, and $a \approx ~20~AU$. With these values the predicted KH growth rate, for planar 
geometry is $5\times 10^{-8}~sec^{-1}$.

To verify that this instability seen in the 3D MHD code is indeed caused by a KH 
type instability arising in the jet-sheet, we studied a much simpler 2D
configuration, analogous to that studied by \cite{einaudi}.
In our case, the radial magnetic field in the flow direction is
negligible, while there is a strong sheared magnetic field 
perpendicular to this direction.  The sheet thickness is much smaller 
than both the sheet length and it's longitudinal extension. Therefore, in 
the simplified geometry \citep{bettarini}, we model this by considering
an initial equilibrium structure in cartesian geometry with velocity $V_{y}$
that varies in the $x$ direction and is uniform in the $y$ and $z$ directions.
The jet-sheet is therefore infinite in the $y-z$ plane. 
The line profiles for the 
density, temperature and magnetic field were extracted from the case 
described in Section 2.1. Figure 7a and 7b presents the initial 
profiles taken for the velocity of the jet $V0_{y}$ as a function of 
latitude $x$ and for the magnetic field $B0_{z}$ as a function of the 
latitude $x$. 

In particular, while in the 3D jet the temperature varies by a factor of 
almost 10, the density within the jet decreases toward the center by 
a factor of 3, apart from a small central spike (and corresponding 
temperature decrease). The initial conditions for the 2D run 
were therefore chosen to be a uniform 
density with a higher temperature in the jet, in pressure equilibrium 
with the transverse magnetic field changing sign in the center of the 
jet.
The code is a 2.5D compressible MHD code which is spectral in one 
direction (the flow direction $y$) and uses compact finite 
differences on a non-uniform grid achieving a spatial resolution of 
$0.07~AU$ in the $x$ direction. The overall number of grid points was 
$130$ in the $y$ direction and $128$ in the $x$ direction. Non-reflective 
boundary conditions are imposed in this direction using the method of 
projected characteristics as in \cite{einaudiII}. Periodic boundary conditions
are imposed on the $y$ direction. (The $x$ and $y$ axes in the 2.5D run correspond, 
respectively, to the $z$ and $x$ axes in the 3D run.)  Initially, the jet-magnetic field 
equilibrium structure is perturbed by random noise. After an initial transient, 
the jet develops a sinuous instability, as may be seen by inspecting 
the density contours shown for different times in Fig. 8. Initially the density is  uniform along the jet (the left
upper panel).  We can see that as the time evolves the jet starts to
oscillate in a sinuous mode. Because the jet in this case is of 
infinite extent, one can not observe the flag-like oscillation  
involving a total departure from the $z=0$ axis as seen in the 3D run.
Figure 9 shows the growth rate measured by \cite{bettarini}.  The
value obtained is $\Gamma=5.0 \times 10^{9}~sec^{-1}$ for the 
maximum wavelength $\lambda = 25~AU$. These values are almost 
identical to what we 
observe in the 3D runs, for the growth rate and for the $\lambda$, 
which reinforces the idea that the jet 
oscillation is due to the development of the KH instability. 

\section{Conclusions and Discussion}
In the present study we increase and extend the spatial resolution obtained in 
\citet{opher} to study in more details the behavior of the jet. 
The jet-sheet structure is again present and extends 
for $150~AU$ beyond the Termination Shock. Turbulent vortices with sizes $\sim 
20~AU$ are present due to the shear between the jet flow and the 
surrounding flow being pushed aside. The jet is unstable as in 
\citet{opher}. We show that the a de Laval nozzle occurs at the TS 
near the current sheet and it is responsible for accelerating the flow 
to supersonic values after the shock. We address how the width of the current sheet 
depends on spatial resolution showing that as the resolution increases, the 
width tend to a finite value of $4~AU$. As we increase the spatial resolution, 
the profiles of the velocity tend to a common shape. This result we believe, indicate that the 
width of the jet in the high resolution calculation, is independent of the grid resolution, and depends on 
physical conditions rather than on numerical resolution.

In order to verify and confirm that the instability of the current 
sheet is a KH type of instability we have also run a 2D compressible MHD case. 
We used a simple geometry of an infinite jet-magnetic field equilibrium using 
the profiles for the magnetic field and velocity from our full 3D 
runs. The jet is unstable and develop a sinuous instability as 
observed in the 3D run. The growth rate is $5.0 \times 10^{-9} 
sec^{-1}$ in excellent agreement with the value measured in the 
3D runs ($5.4\times 10^{-9}~sec^{-1}$). The growth rate had a maximum 
for $\lambda =25~AU$ again in agreement with the 
$\lambda$ measured in the 3D runs, $23~AU$. This reinforce the idea 
that the jet oscillation is due to a KH instability. 

The presence of the extended jet, turbulence and the 
magnetohydrodynamic instabilities reinforce the view that the boundary between the 
solar wind and the interstellar wind is a dynamic environment. It is highly distorted 
discontinuity and that magnetic effects are a major player in this 
region. Still other important effects such as neutrals and the tilt of the 
magnetic axis in respect to the rotation axis need to be included 
before we can address definitively how this boundary appears. Also, when the interstellar magnetic 
field is included, reconnection between the solar and interstellar magnetic field will occur.

The inclusion of a tilted heliospheric current sheet very likely will introduce 
qualitative changes in the picture represented in this paper. \citet{nerney1} investigated analytically the solar
cycle imprint on the Heliosheath. They predicted that the 11-year solar cycle would imprint an
alternating magnetic polarity envelopes in the Heliosheath.
On a much finer scale, there will be strongly mixed polarity regions between the magnetic envelopes due to
the reverse magnetic field polarity due to the 25.5-day solar rotation period.
The velocity shear instability requires hundreds of years to develop,
while the solar magnetic field changes polarity in much shorter time scales.
Thus, in order to have a complete picture of the heliosheath, time-dependent boundary conditions 
on the solar magnetic field need to be included.
By the picture described in this paper, we expect however, that this region will be very turbulent
and dynamic.

The {\it Voyager} spacecraft is now at $\approx 85~AU$ and is approaching the Termination Shock 
(there is current controversy on whether it already encounter it \citep{krimi,mcdonald}). {\it Voyager} travels 35${}^{o}$ 
above the ecliptic plane towards the nose of the heliosphere. We expect that {\it Voyager} will encounter an extremely 
turbulent dynamic region that will have a major effect in the energetic 
particle detected by the instruments on {\it Voyager}.

Finally, for other magnetized rotating stars in motion with respect 
to the interstellar medium we expect to find similar phenomena. We 
expect the presence of an extended jet or disk with length of $\sim 
150~AU$ and a width of $~10~AU$ beyond the location where stellar 
winds are shocked in respect to the interstellar medium that 
surround them.

\acknowledgments
This work is a result of the research performed at the Jet Propulsion 
Laboratory of the California Institute of Technology under a contract 
with NASA. The University of Michigan work was also supported by 
NASA. GT was partially supported by the Hungarian Science Foundation 
(OTKA, grant No. T037548). We thank F. Rappazzo and S. Landi for 
useful discussions.

\appendix
  \begin{center}
    {\bf APPENDIX: DEPENDENCE OF THE WIDTH OF THE JET ON SPATIAL RESOLUTION} 
  \end{center}
In this appendix we present the grids for the three cases. Figure 10-12 show the 
contours of the azimuthal magnetic field $B_{y}$ for the coarse, refined and super-refined-extended cases. 
The red lines indicate the change in the grid resolution. The numbers indicate the 
resolution in each region, in $AU$. For the coarse grid, the resolution between the TS and the 
HP, at the HCS, is $6.0~AU$. For the refined grid, the resolution between the TS and HP, at the 
HCS, is $1.5~AU$. In the super-refined-extended case the resolution between the TS and 
HP, at the HCS, is $0.75~AU$.

Figure 13 shows the magnetic field (solid line) and the velocity profiles 
(dashed line) for the super-refined extended case. Notice that the 
magnetic field profile is inverted. We can see how the flow profile 
adjust itself to the magnetic one. For the three cases the flow profile is broader than 
the magnetic one. As we increase the resolution, the two profiles almost overlap, and 
the width of the jet is shaped by the HCS width. 

\vskip0.2in


\begin{figure*}[ht!]
\begin{minipage}[t] {0.5\linewidth}
\begin{center}
\includegraphics[angle=0,scale=0.3]{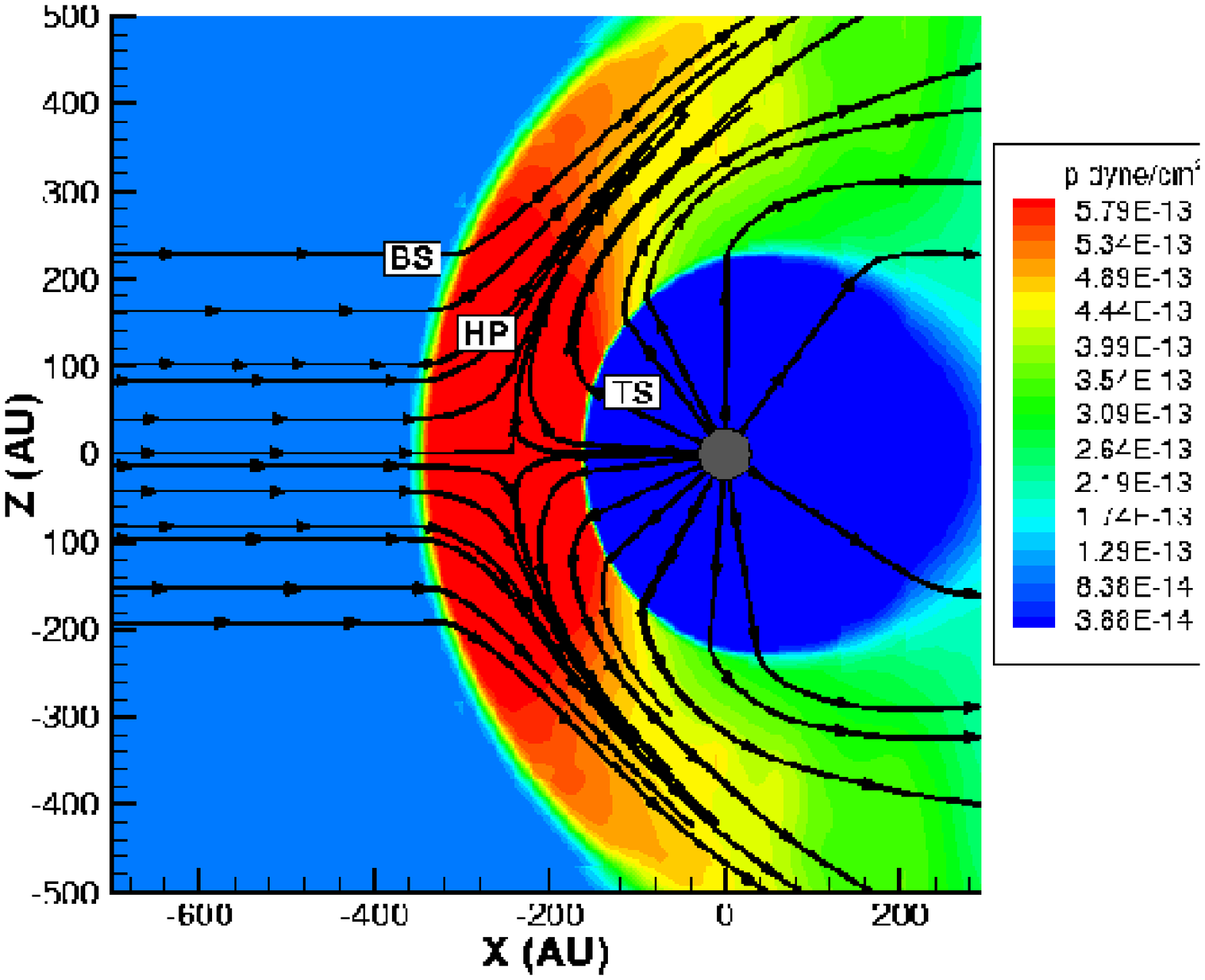}
\end{center}

\end{minipage} \hfill

\begin{minipage}[t] {.5\linewidth}
\begin{center}
\includegraphics[angle=0,scale=0.3]{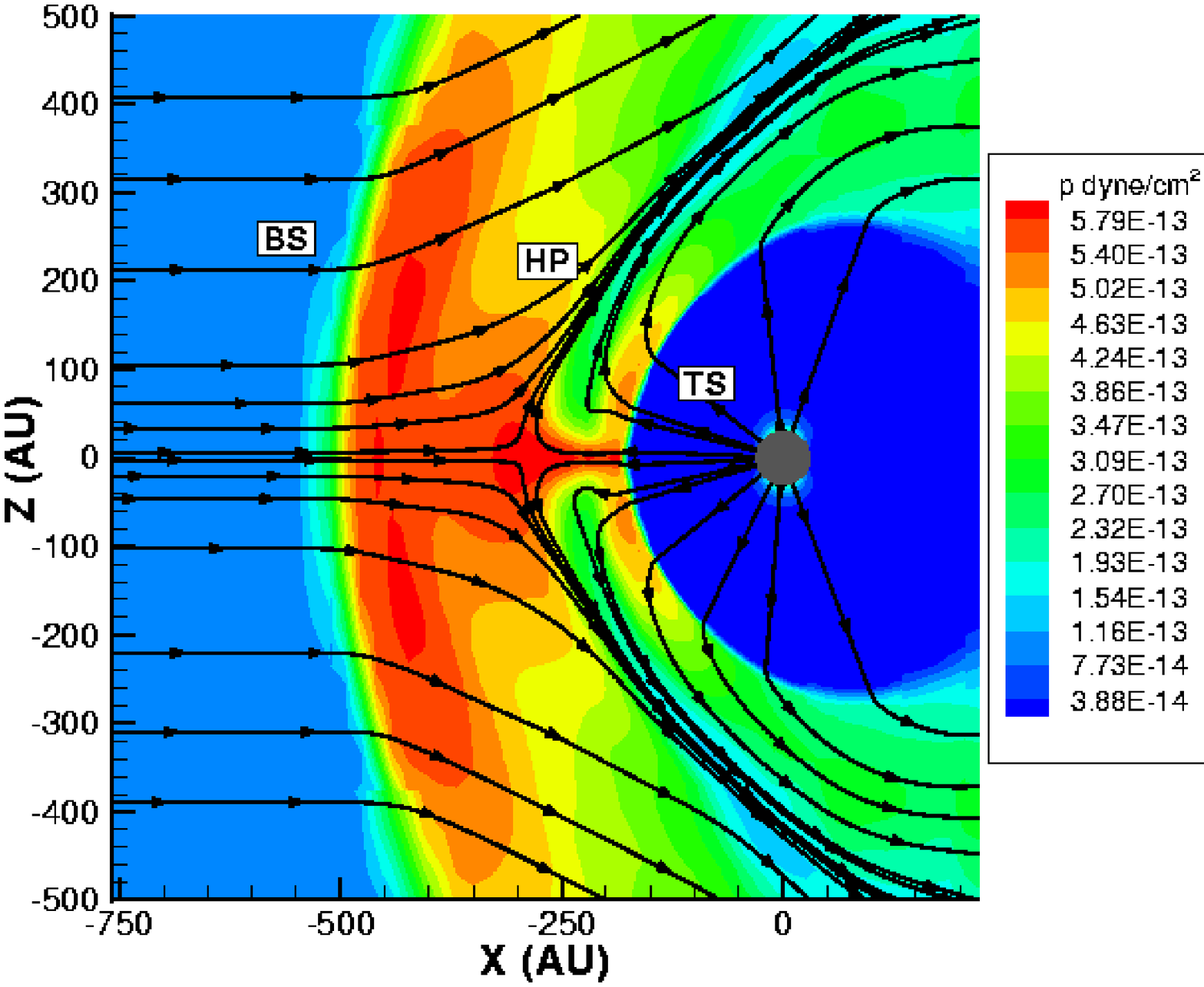}
\end{center}

\end{minipage}\vfill

\caption{(a) Contours of pressure at $t=1.0 \times 10^{9}~sec$ in 
the meridional plane (x-z) without the solar magnetic field. Same as (a) but with the solar magnetic field. The case presented here is the {\it 
coarse} case, with resolution of the grid at the current sheet of 
$\sim 3AU$.}
\label{fig1}
\end{figure*}

\begin{figure*}[ht!]

\begin{minipage}[t] {0.5\linewidth}

\begin{center}

\includegraphics[angle=0,scale=0.3]{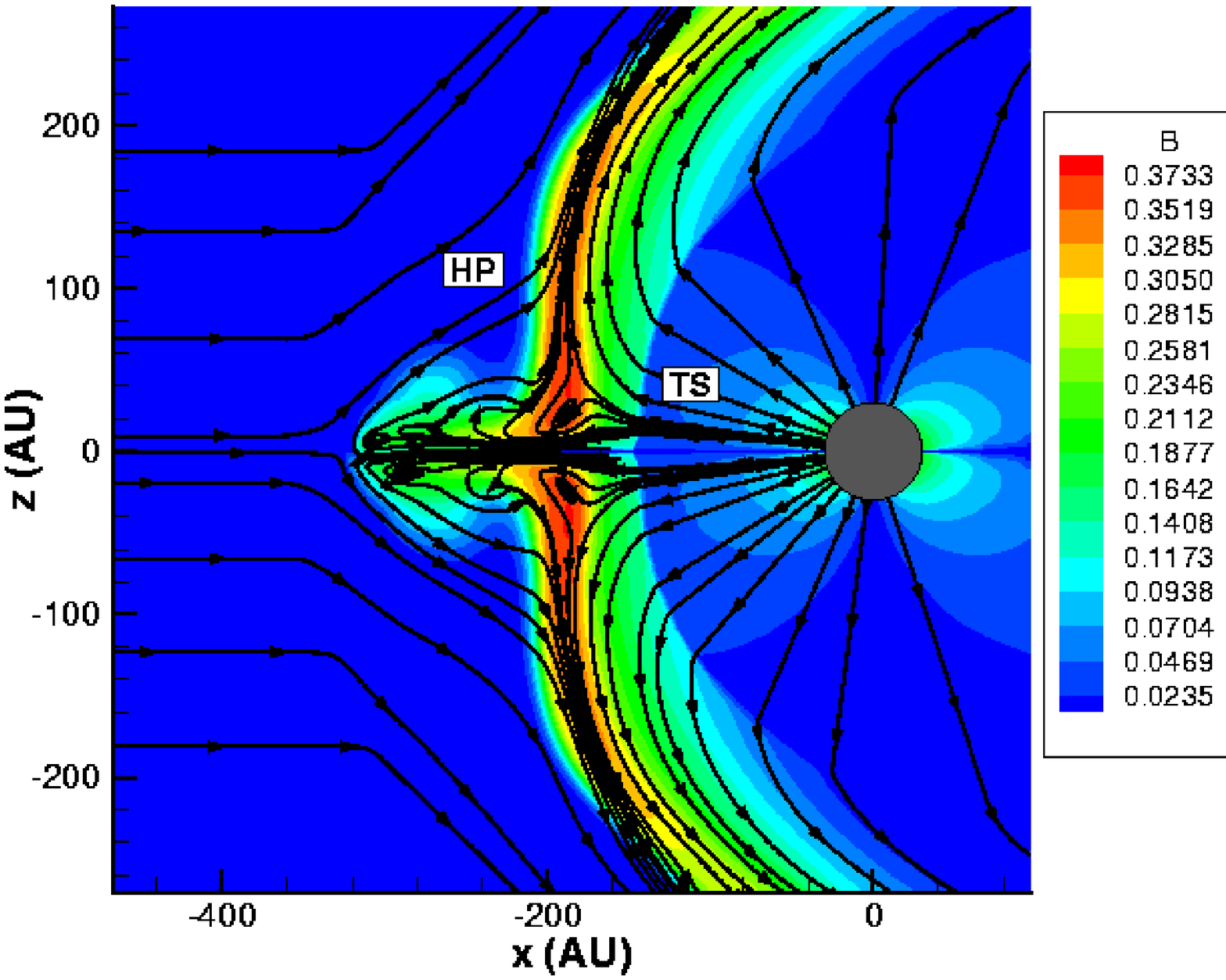}

\end{center}

\end{minipage} \hfill

\begin{minipage}[t] {.5\linewidth}

\begin{center}

\includegraphics[angle=0,scale=0.3]{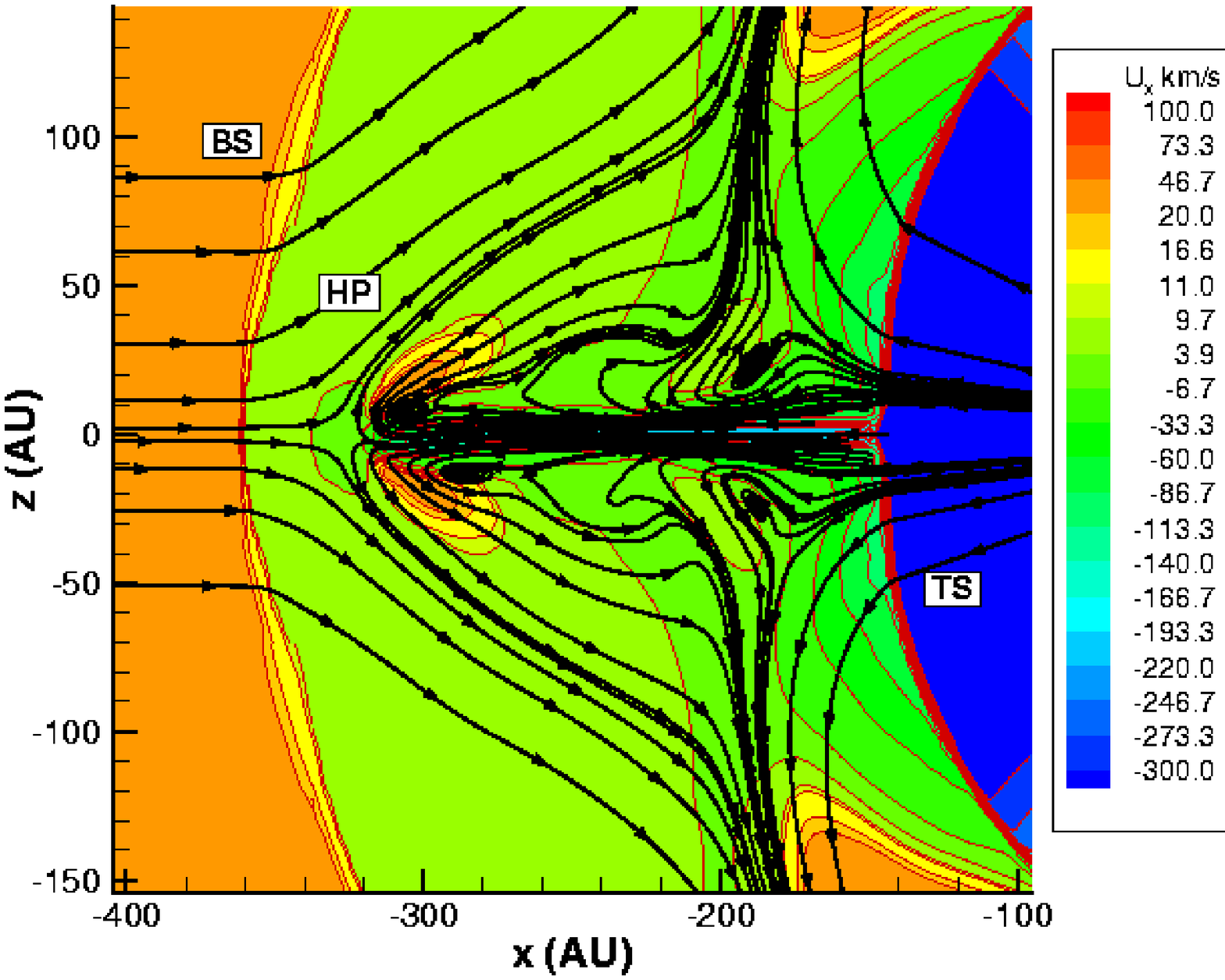}

\end{center}

\end{minipage}\hfill

\begin{minipage}[t] {0.5\linewidth}

\begin{center}

\includegraphics[angle=0,scale=0.3]{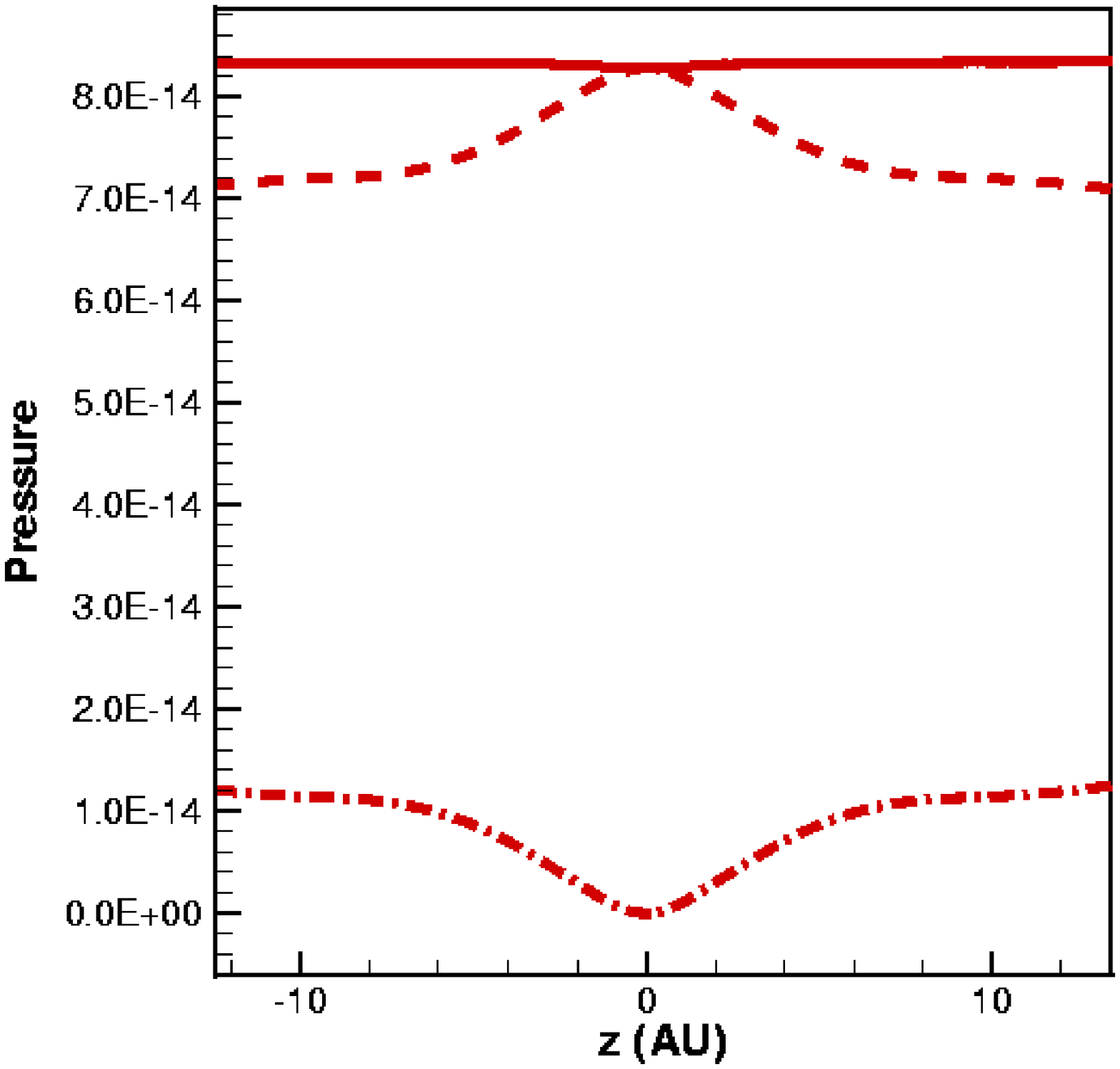}

\end{center}

\end{minipage} \hfill

\begin{minipage}[t] {0.5\linewidth}

\begin{center}

\includegraphics[angle=0,scale=0.3]{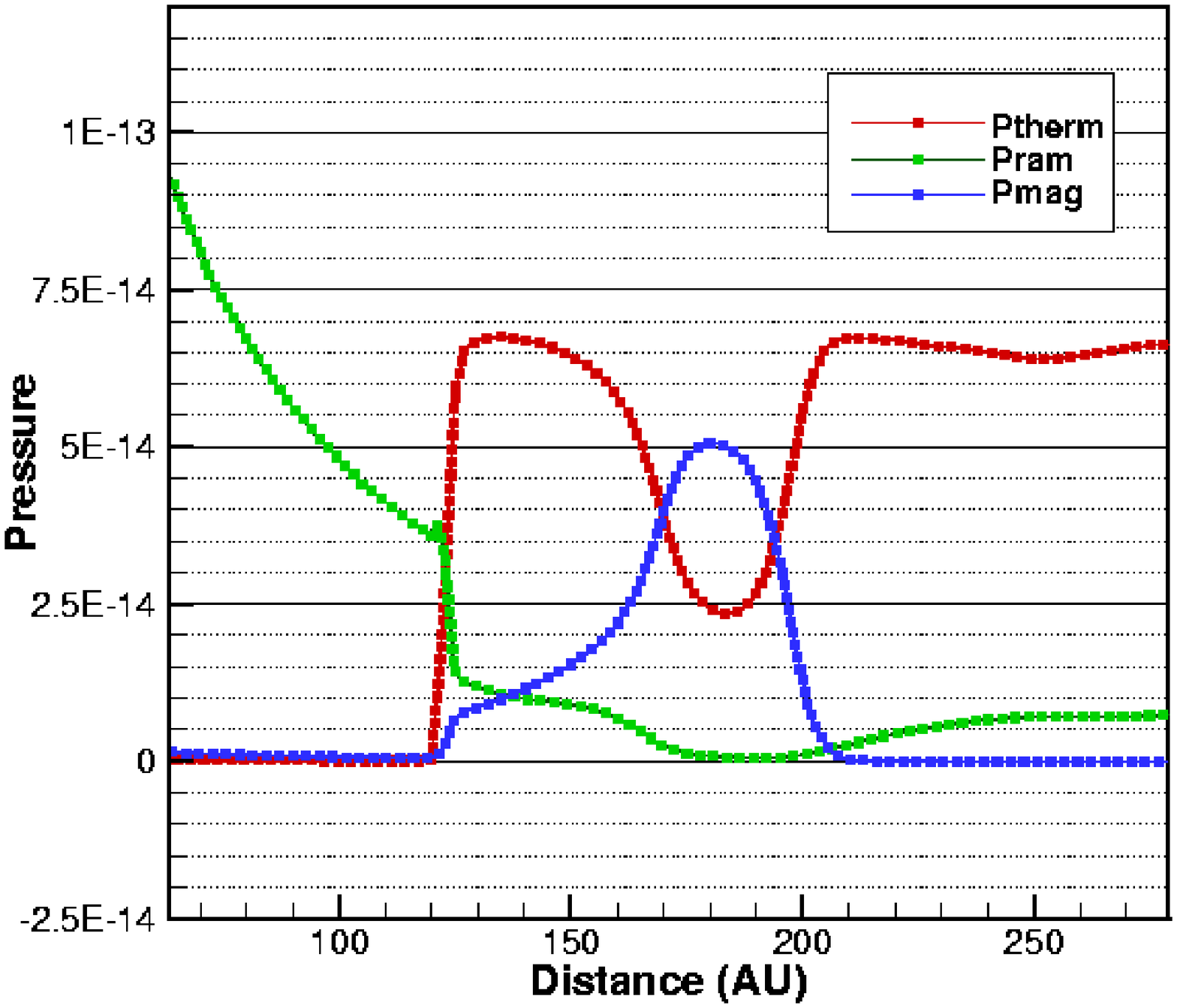}

\end{center}

\end{minipage} \vfill

\caption{(a) Contours of magnetic field at $t=1.3 \times 10^{9} 
sec$ (scale ranging from $0-0.36nT$) in the meridional plane (x-z). 
The red lines indicate 
the boundaries when the grid refinement changes. The HP, TS and the 
heliocurrent sheet (HCS) are indicated. Also are shown the flow 
streamlines (black lines). Note the solar dipolar magnetic field. The red regions 
are the ``magnetic ridges''. (b) Same as Fig 1a but contours of $Ux$. We can see the jet extending 
for $150~AU$ beyond the TS, to almost touching the BS. Also, the 
presence of the back flow sweeped aside by the jet. (c) Vertical 
line cut at $x=161~AU$ (downstream 
the TS). The solid curve is 
the total pressure, the dashed ones is the thermal pressure and the 
dash-dot-dash one is the magnetic pressure. (d) Line cut $60^{\circ}$ 
above the equator showing the thermal pressure $P_{therm}$, the ram 
pressure $P_{ram}$ and the magnetic pressure $P_{mag}$.}
\label{fig2}
\end{figure*}

\begin{figure*}
\begin{center}
\includegraphics[angle=0,scale=0.4]{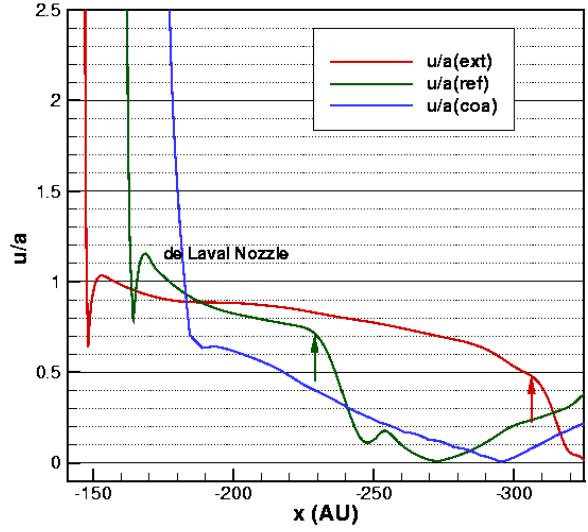}
\end{center}
\caption{Line plot of the equatorial cut at $t=1.5\times 10^{9} sec$ 
of the ratio velocity per sounds speed $a$ vs. $x$. Three cases are
shown: super-refined-extended (red), refined (green) and coarse 
(blue). The green and blue refers to the cases at Opher et al. 2003. 
As we increase the resolution the jet(red) extends farther away. The 
de Laval nozzle is still present. The refined jet (green) lose its 
power at $x=-230~AU$ (indicated by the green arrow), while the change 
in resolution from $1.5$ to $3.0~AU$ occurs at $x=-320~AU$. The 
extended jet (red) lose its power at $x=-320~AU$ (indicated by the 
red arrow) while the change in resolution (from $0.75$ to $1.5~AU$ at 
$x=-350~AU$ and from $1.5$ to $3.0~AU$ at $x=-365~AU$.} 
\label{fig3}
\end{figure*} 

\begin{figure*}[ht!]
\begin{minipage}[t] {0.5\linewidth}
\begin{center}
\includegraphics[angle=0,scale=0.3]{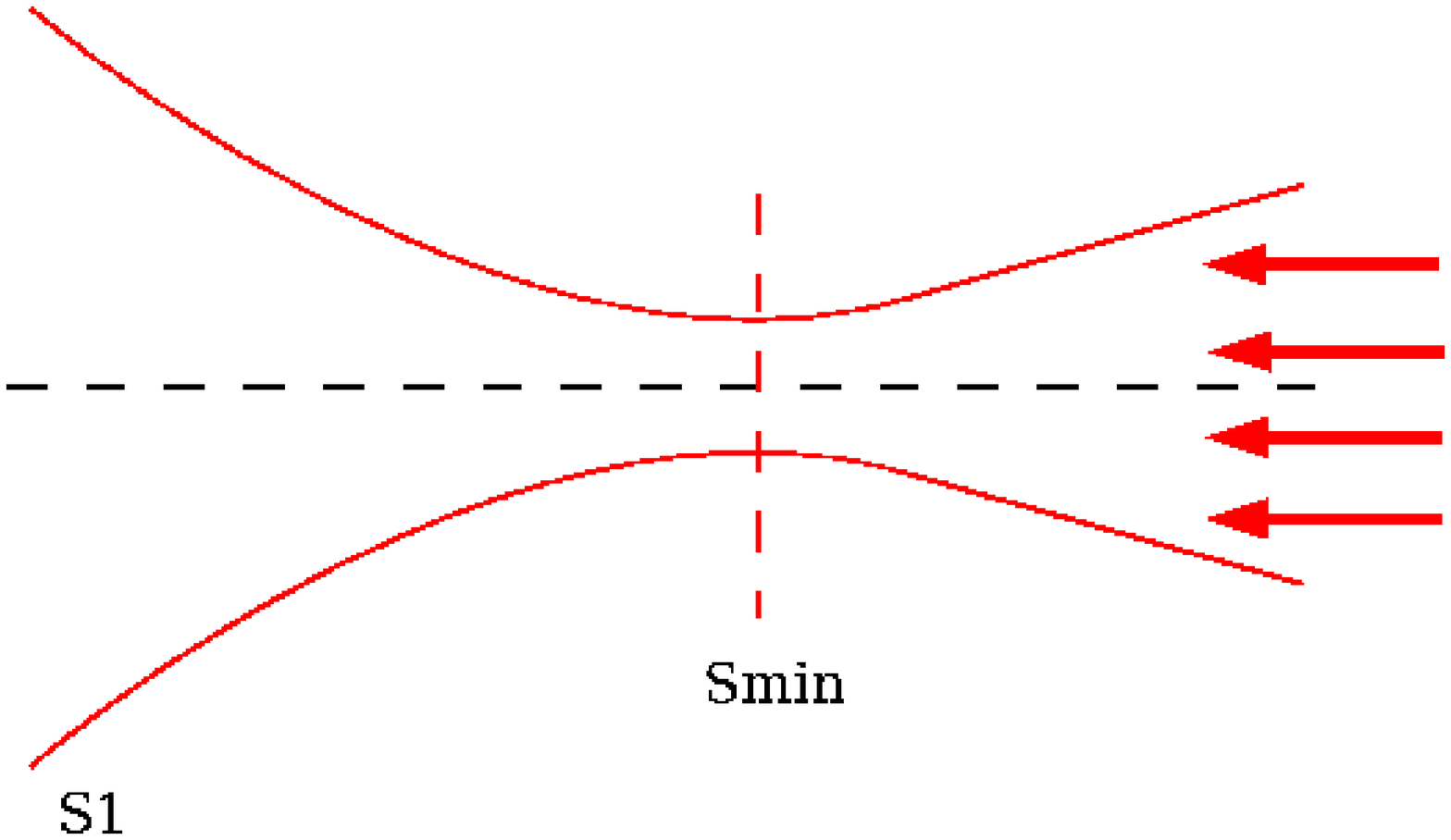}
\end{center}
\end{minipage} \hfill
\begin{minipage}[t] {.5\linewidth}
\begin{center}
\includegraphics[angle=0,scale=0.3]{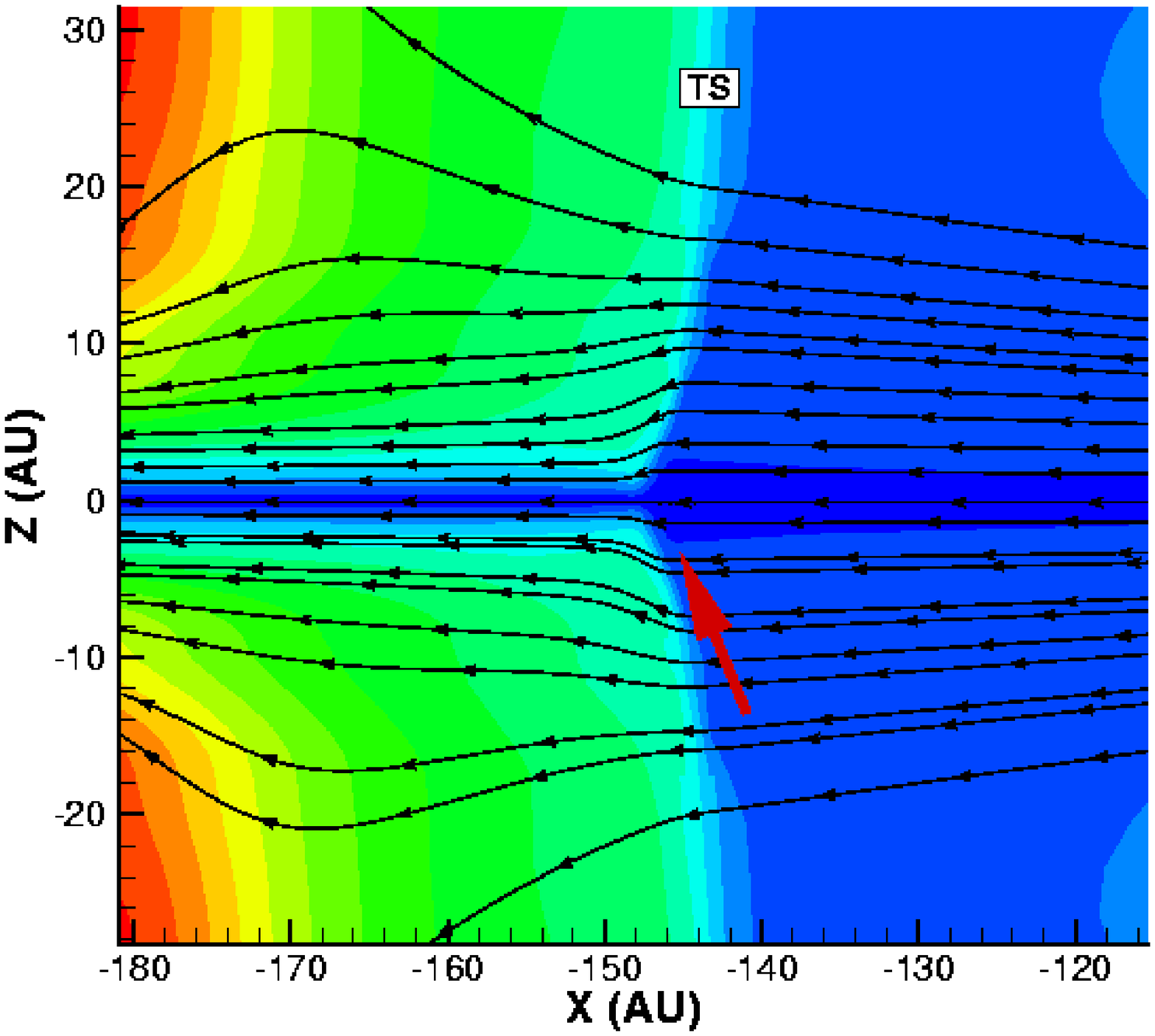}
\end{center}
\end{minipage}\vfill
\caption{(a) Schematic drawing illustrating the formation of the 
{\it de Laval Nozzle}. (b) Close view of the region of the {\it de 
Laval} nozzle just after the TS. The red arrow indicate the de Laval 
nozzle, where the flow converges.}
\label{fig4}
\end{figure*}

\begin{figure*}[ht!]
\begin{minipage}[t] {0.5\linewidth}
\begin{center}
\includegraphics[angle=0,scale=0.3]{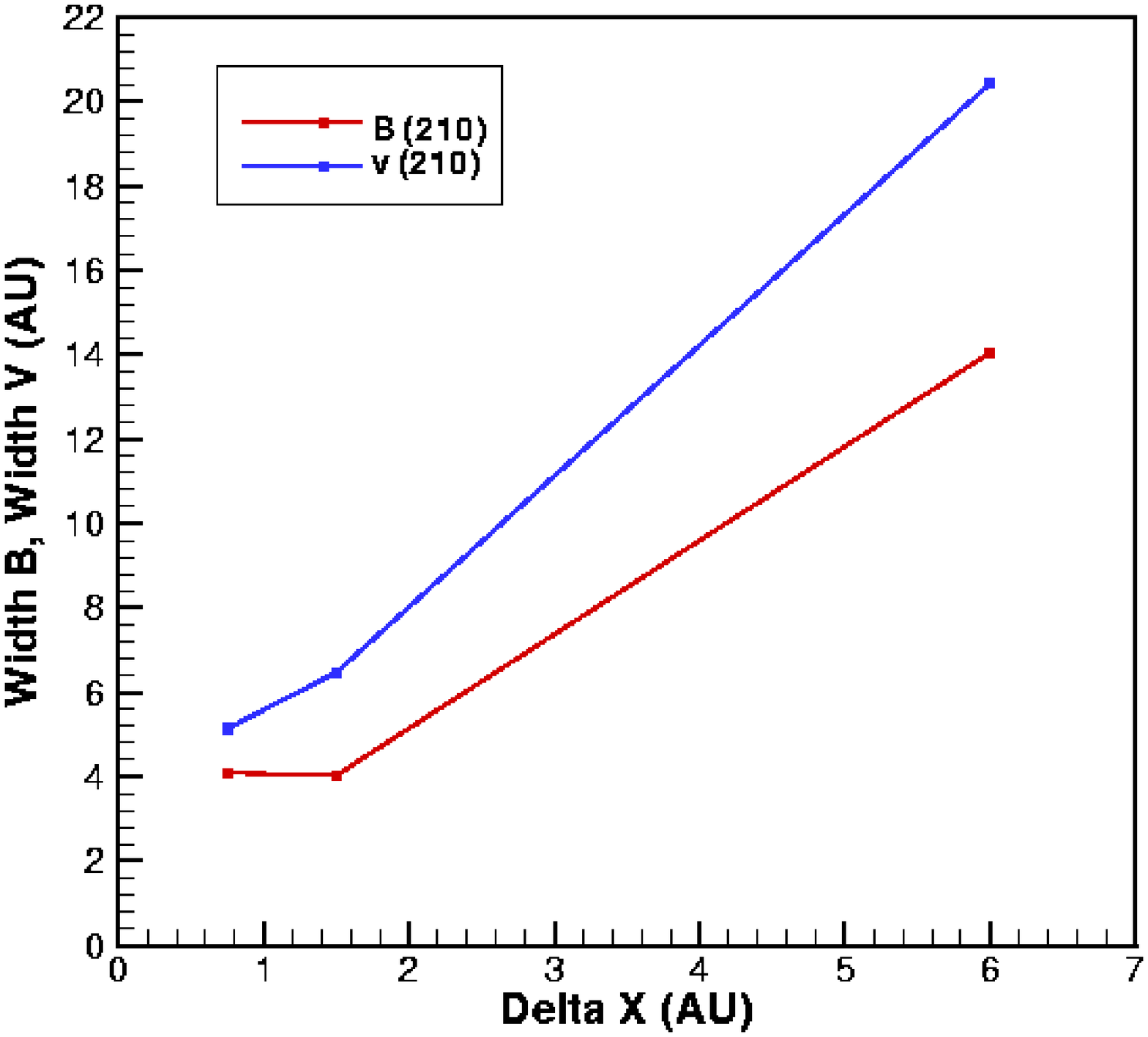}
\end{center}
\end{minipage} \hfill

\begin{minipage}[t] {.5\linewidth}
\begin{center}
\includegraphics[angle=0,scale=0.3]{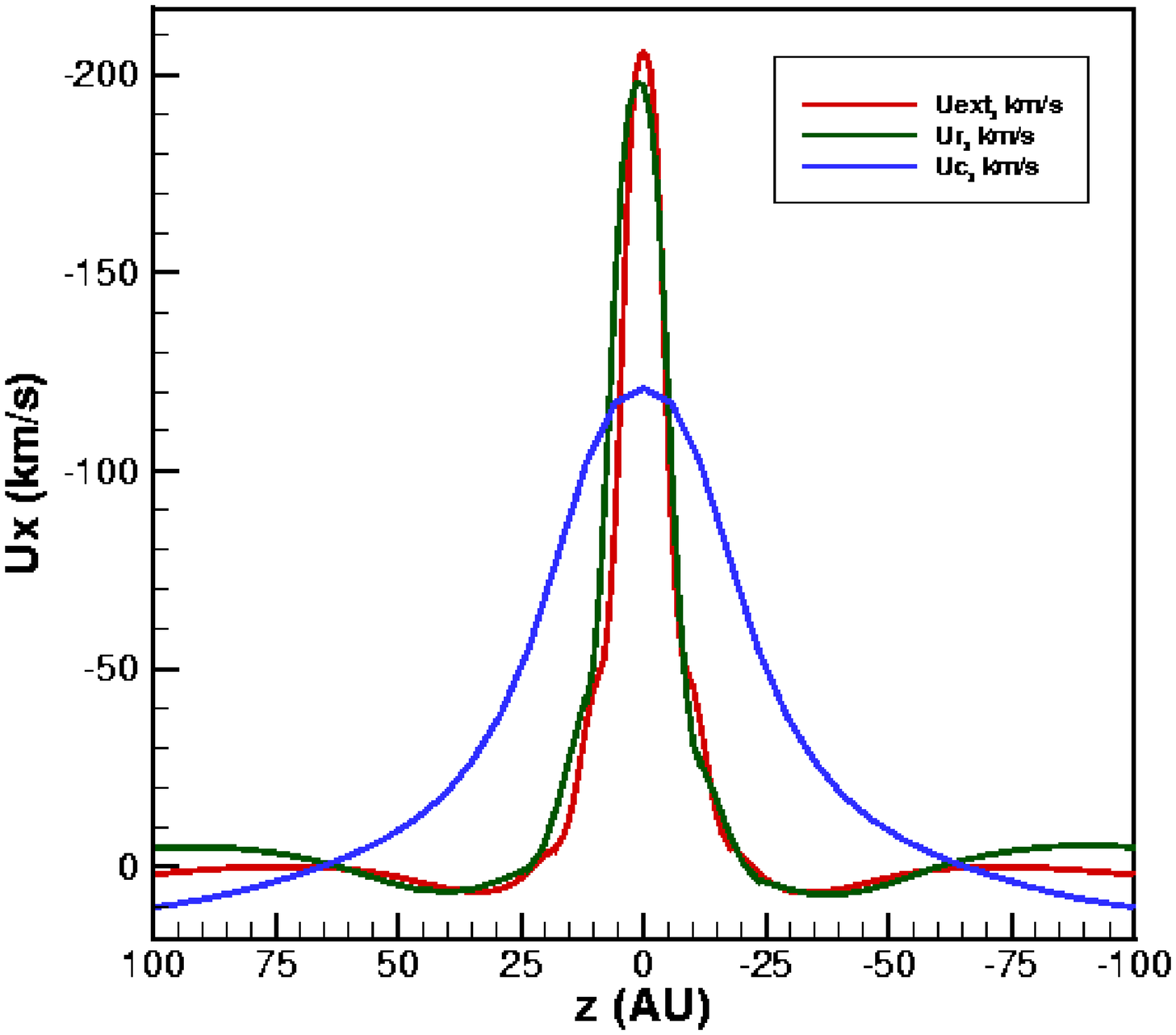}
\end{center}
\end{minipage}\vfill
\caption{(a) Width of the current sheet and the jet vs resolution 
measured at $x=-210AU$. The red line is the magnetic field width and the 
blue line is the velocity width} (b) Velocity profiles for the coarse, 
refined and super-refined-extended cases.
\label{fig5}
\end{figure*}

\begin{figure*}[ht!]
\begin{minipage}[t] {0.5\linewidth}
\begin{center}
\includegraphics[angle=0,scale=0.3]{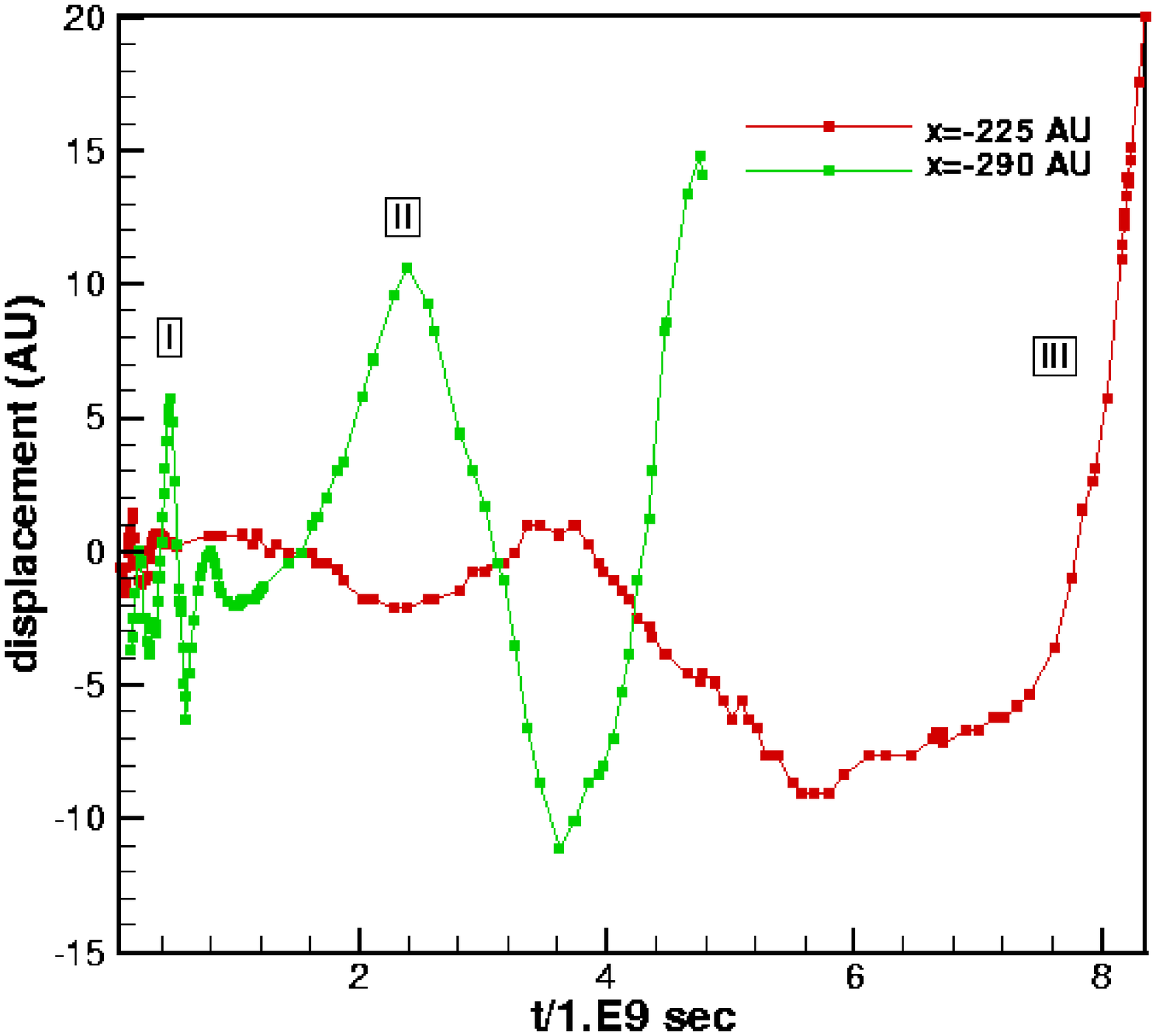}
\end{center}
\end{minipage} \hfill
\begin{minipage}[t] {.5\linewidth}
\begin{center}
\includegraphics[angle=0,scale=0.3]{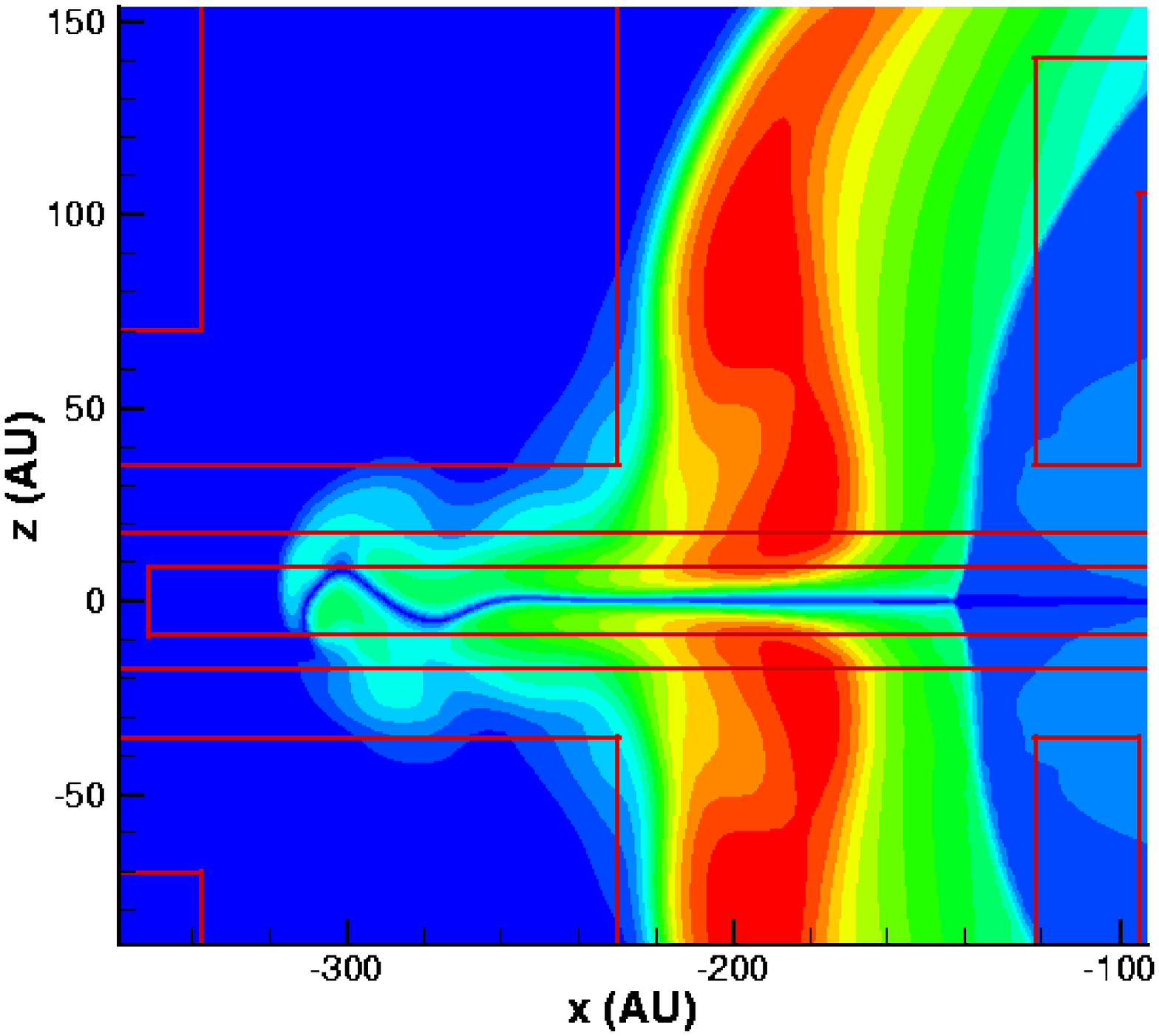}
\end{center}
\end{minipage}\vfill
\caption{(a) Displacement of the current sheet (AU) vs $t/10^{9} 
sec$ for $x=-225AU$ (red) and $x=-290AU$ (green). Indexes (I)-(III) 
marks the different regions of time scales of oscillation. (b) 
Contours of the magnetic field at $t=16.8~years$. The scale is the 
same as Fig2a.}
\label{fig6}
\end{figure*}

\begin{figure*}
\begin{minipage}[t] {0.5\linewidth}
\begin{center}
\includegraphics[angle=0,scale=0.3]{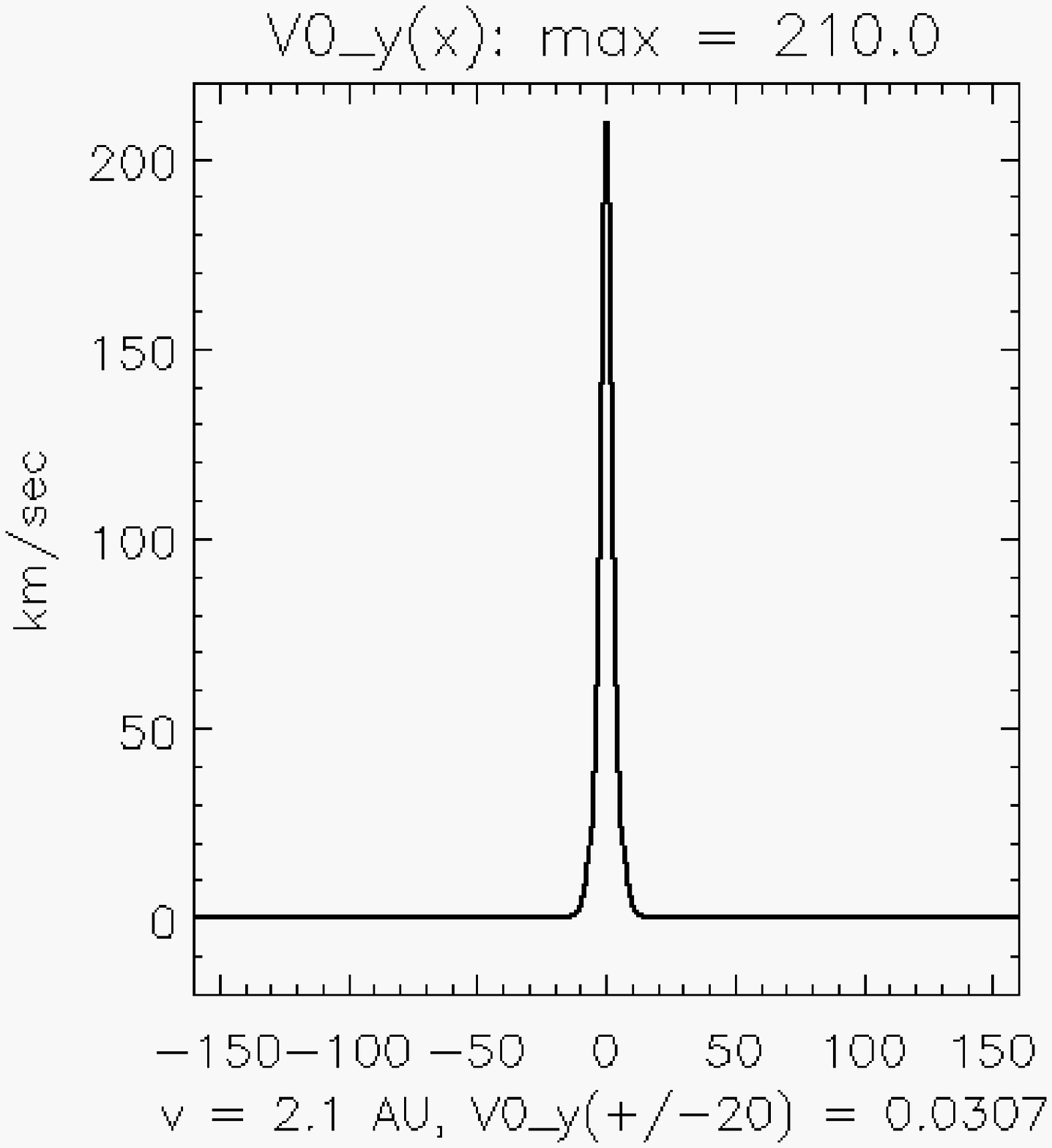}
\end{center}
\end{minipage} \hfill
\begin{minipage}[t] {.5\linewidth}
\begin{center}
\includegraphics[angle=0,scale=0.3]{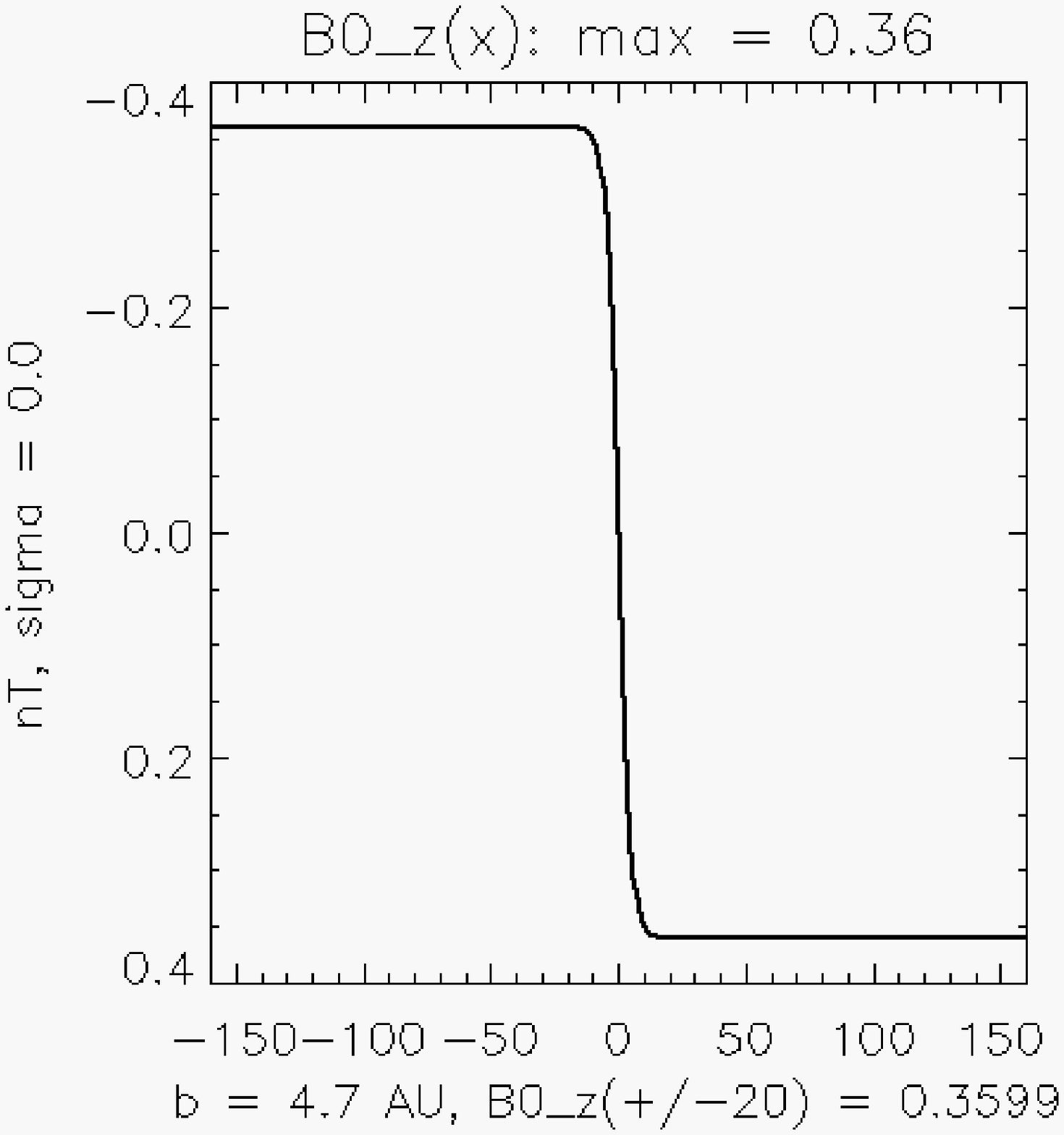}
\end{center}
\end{minipage}\vfill
\caption{(a) Initial profile for the velocity of the jet $V0_{y}$ 
as a function of latitude $x$ for the 2D simulations. (b) Initial profile for the 
magnetic field $B0_{z}$ as a function of the latitude $x$.}
\label{fig7}
\end{figure*}

\begin{figure*}
\begin{center}
\includegraphics[angle=0,scale=0.3]{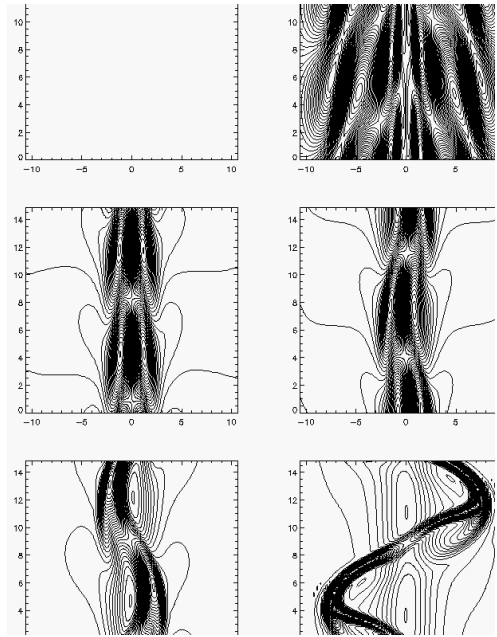}
\end{center}
\caption{Density contours for the 2D compressible MHD run in the $xy$ coordinate system.} 
\label{fig8}
\end{figure*}

\begin{figure*}
\begin{center}
\includegraphics[angle=0,scale=0.4]{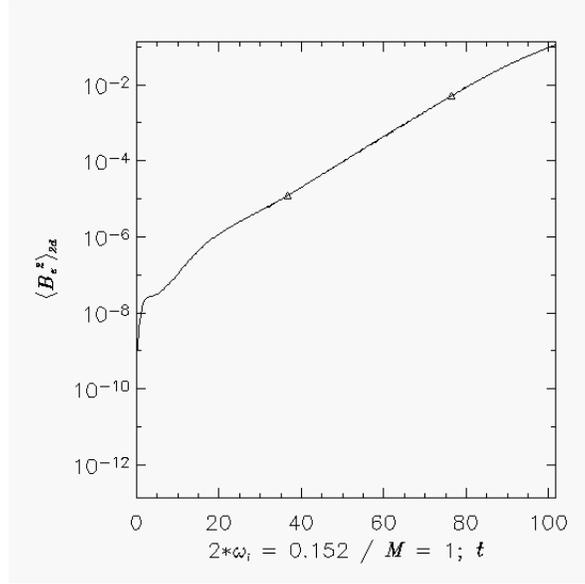}
\end{center}
\caption{Displacement of the current sheet vs $t$ for the 2.5D compressible MHD run.} 
\label{fig9}
\end{figure*}

\begin{figure*}
\begin{center}
\includegraphics[angle=0,scale=0.4]{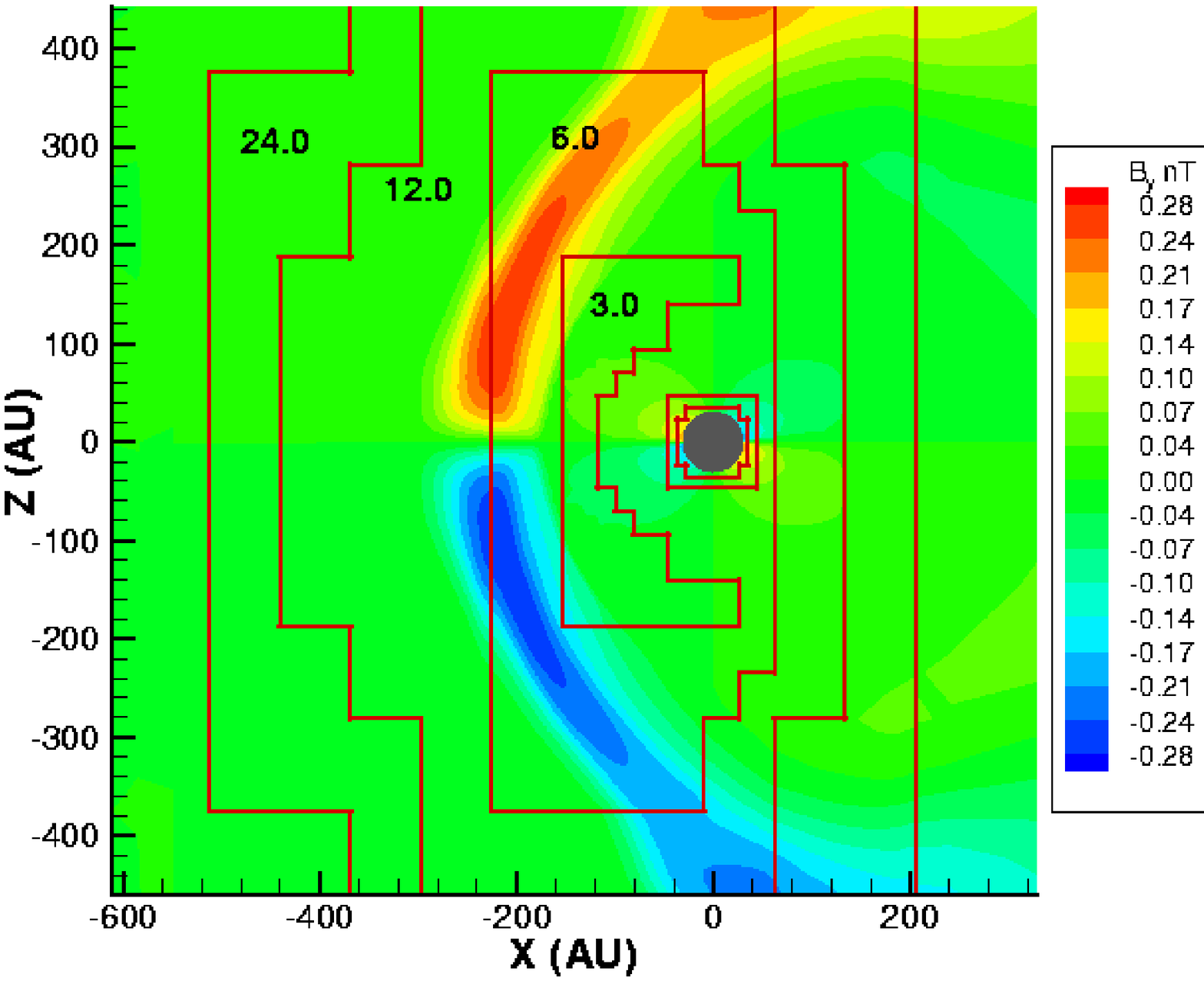}
\end{center}
\caption{The contours of the azimuthal magnetic field, $B_{y}$, for $t=10^{9}~sec$ for the coarse case. The red lines indicate 
the changes in grid resolution. The numbers indicate the grid resolution (in $AU$).} 
\label{fig10}
\end{figure*} 

\begin{figure*}
\begin{center}
\includegraphics[angle=0,scale=0.4]{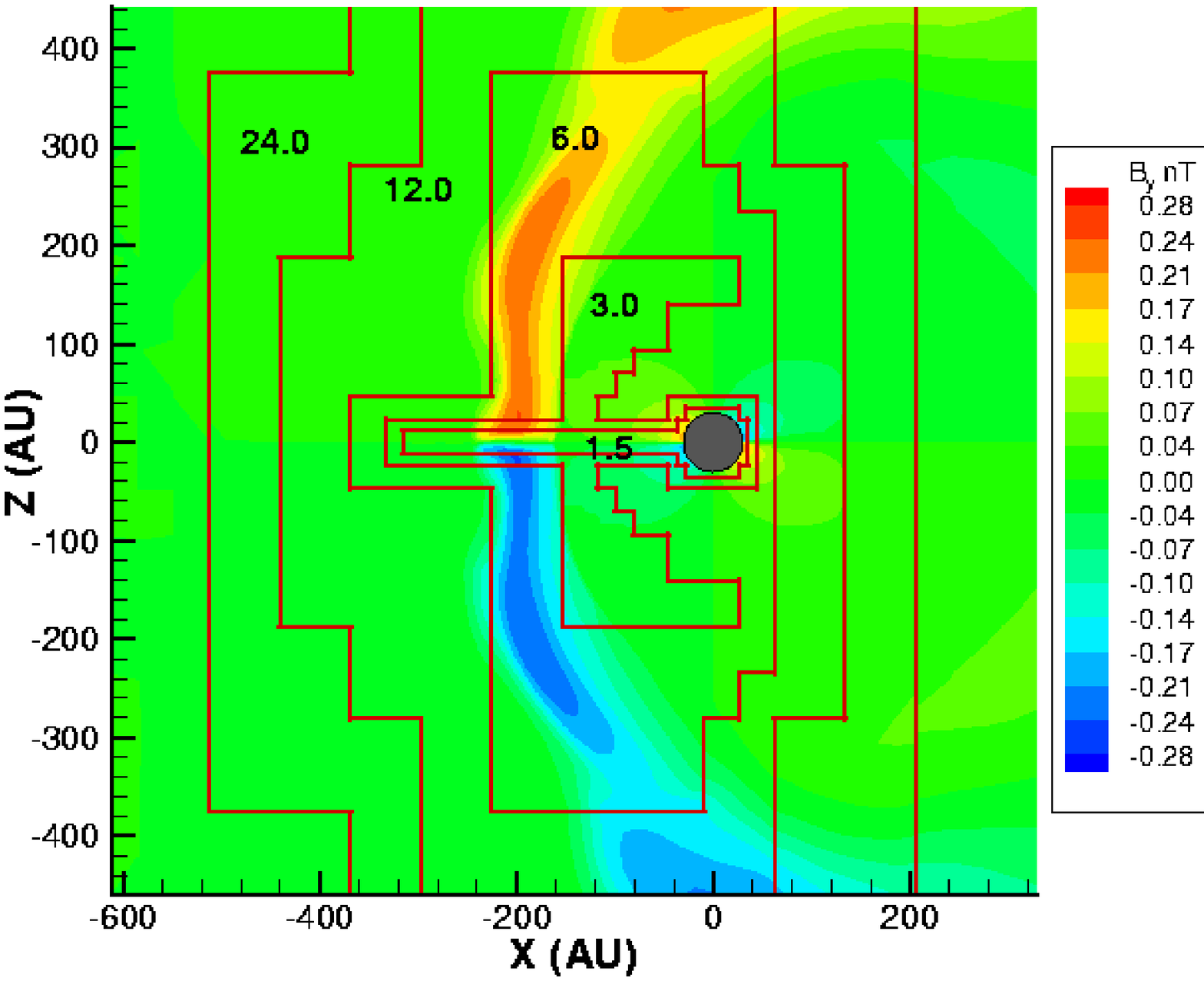}
\end{center}
\caption{The contours of the azimuthal magnetic field, $B_{y}$, for $t=10^{9}~sec$ for the refined case. The red lines 
indicate the changes in grid resolution. The numbers indicate the grid resolution (in $AU$).} 
\label{fig11}
\end{figure*} 

\begin{figure*}
\begin{center}
\includegraphics[angle=0,scale=0.4]{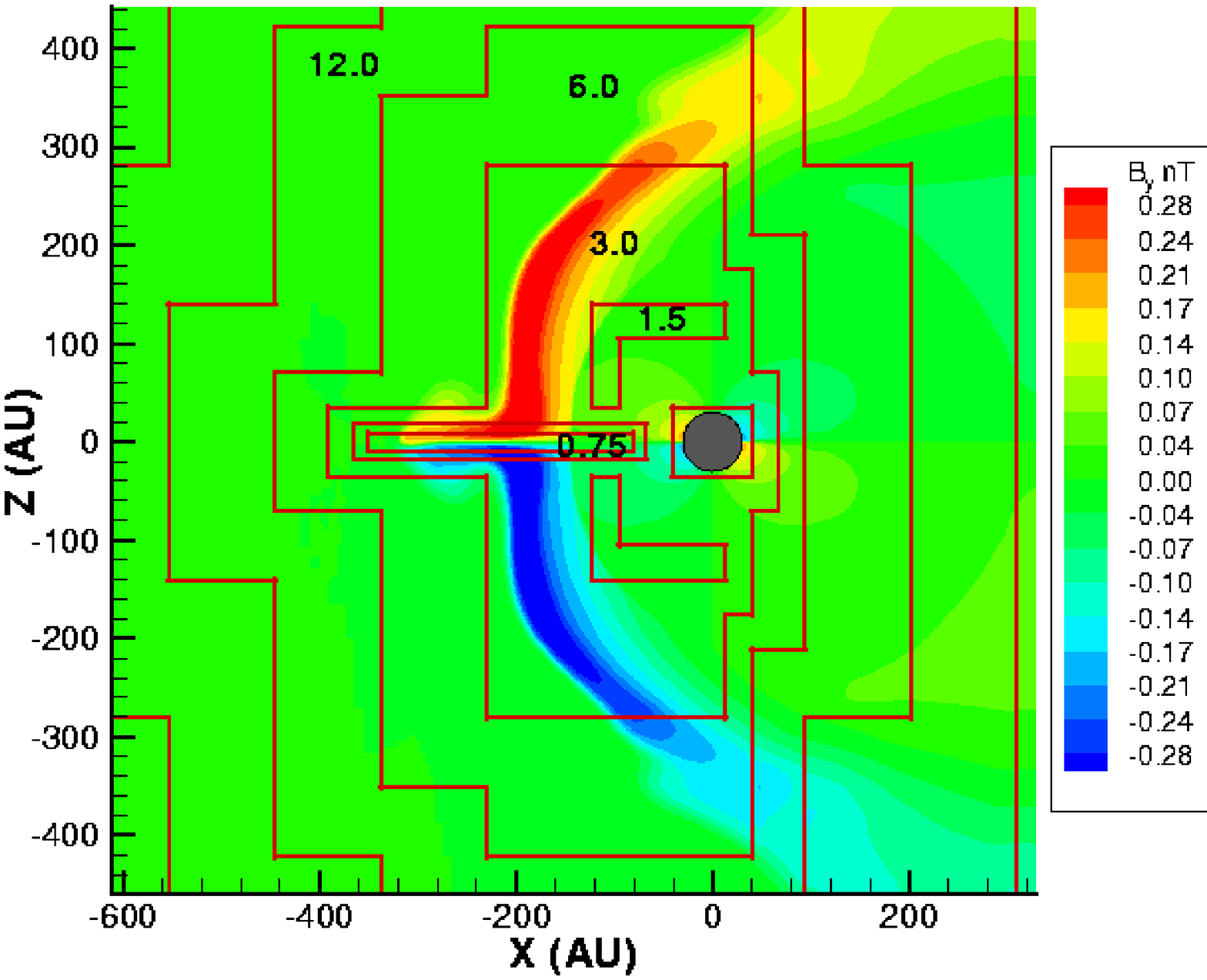}
\end{center}
\caption{The contours of the azimuthal magnetic field, $B_{y}$, for $t=10^{9}~sec$ for the super-refined-extended case. 
The red lines indicate the changes in grid resolution.The numbers indicate the grid resolution (in $AU$).} 
\label{fig12}
\end{figure*} 

\begin{figure*}
\begin{center}
\includegraphics[angle=0,scale=0.4]{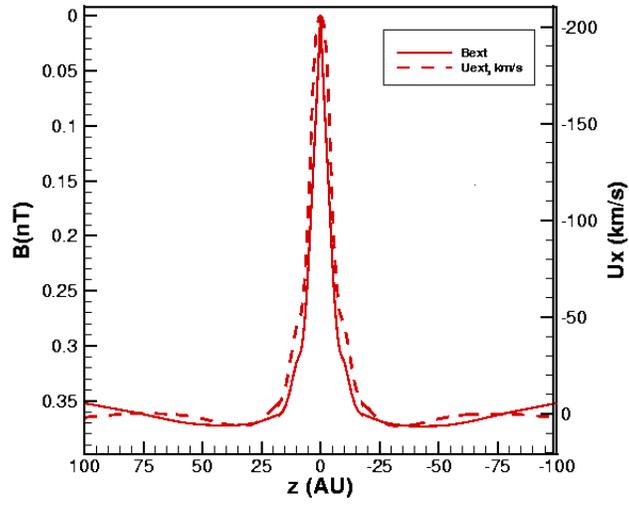}
\end{center}
\caption{Velocity and magnetic field profiles for the super-refined-extended case. The full line
is the magnetic field and the dashed line is the velocity profile. Note that the magnetic field profile is 
inverted.}
\label{fig13}
\end{figure*}


\noindent


\begin{thebibliography}{}

\bibitem[Baranov \& Malama(1993)]{baranov93} Baranov, V. B., \& 
Malama, Y.G. 1993, \jgr, 98, 15157.
\bibitem[Baranov \& Malama(1995)]{baranov95} Baranov, V. B., \& 
Malama, Y.G. 1995, \jgr, 100, 14755.
\bibitem[Baranov \& Zaitsev(1995)]{zaitsev} Baranov, V. B., \& 
Zaitsev, N. A. 1995, Astron. Astrophys. 304, 631.
\bibitem[Baranov, Izmodenov \& Malama(1998)]{baranov98} Baranov, V. 
B., Izmodenov, V. V., \& Malama, Y. G. 1998, \jgr, 103, 9575.
\bibitem[Bettarini(2003)]{bettarini} Bettarini, L. 2003, University of Florence, M.A. Thesis.
\bibitem[Chandrasekhar(1961)]{chandra} Chandrasekhar, S., 1961, {\it Hydrodynamic and Hydromagnetic
Stability}, (Clarendon, Oxford, England).
\bibitem[Dahlburg, Boncinelli \& Einaudi(1998)] {dahlburg} Dahlburg, 
R.B., Boncinelli, P., \& Einaudi, G. 1998, Phys. Plasmas, {\bf 5}, 79.
\bibitem[Dahlburg, Keppens \& Einaudi(2001)]{keppens} Dahlburg, R.B., 
Keppens, R., Einaudi, G. 2001, Physics.Plasmas, 8, 2001.
\bibitem[Einaudi(1999)] {einaudi} Einaudi, G., 1999, Phys. Control. 
Fusion, A293, 41.
\bibitem[Einaudi et al.(2001)] {einaudiII} Einaudi, G., Chibbaro, S., 
Dahlburg, R. B., \& Velli, M. 2001, \apj, 547, 1167.
\bibitem[Frisch(1996)]{frisch} Frisch, P. C. 1996, Space Sci. Rev., 
78, 213.
\bibitem[Izmodenov, Gruntman \& Malama(2001)]{izmodenov} Izmodenov, 
V., Gruntman, M., \& Malama, Y. G. 2001, \jgr, 106, 10681.
\bibitem[Krimigis et al.(2003)]{krimi} Krimigis, S.M., Decker, R.B., Hill, M.E., 
Armstrong, T.P., Gloecker, G., Hamilton, D.C., Lanzerotti, L.J., \& Roelof, E.C. 2003, 
Nature, 426, 45.
\bibitem[Landau \& Lifshitz(1987)]{landau} Landau, L.D., \& Lifshitz, 
E. M. 1987, Fluid Mechanics (Oxford: Pergamon)
\bibitem[Liewer et al.(1996)] {liewer} Liewer, P.C., Karmesin, S.R., 
\& Brackbill, J.U. 1996 \jgr, 101, 17119.
\bibitem[Karmesin et al.(1995)] {liewer2} Karmesin, S.R., Liewer, P.C., \& 
Brackbill, J.U. 1995 \grl, 22, 1153.
\bibitem[Linde et al.(1998)]{linde} Linde, T.J., Gombosi, T.I., Roe, 
P.L., Powell, K.G., \& DeZeeuw, D.L. 1998, \jgr, 103, 1889.
\bibitem[McDonald et al.(2003)]{mcdonald} McDonald, F.B., Stone, E.C., Cummings, A.C., 
Heikkila, B., Lai, N., \& Webber, W.R. 2003, Nature, 426, 48.
\bibitem[McNutt, Lyon \& Goodrich(1998)]{mcnutt} McNutt Jr., R.L., 
Lyon, J., \& Goodrich, C. C. 1998, \jgr, 103, 1905.
\bibitem[Muller, Zank \& Lipatov(2000)]{muller} Muller, H-R, Zank, 
G.P., \& Lipatov, A.S. 2000, \jgr, 105, 27419.
\bibitem[Nerney, Suess \& Schmahl(1991)]{nerney} Nerney, S., Suess, 
S.T., \& Schmahl, E.J. 1991, AA, 250,556.
\bibitem[Nerney, Suess \& Schmahl(1995)]{nerney1} Nerney, S., Suess, S.T., \& 
Schmahl, E.J. 1995, \jgr, 100, 3463.
\bibitem[Opher et al.(2003)] {opher} Opher, M. Liewer, P. C.,  
Gombosi, T. I., Manchester, W., DeZeeuw, D. L., Sokolov, I., \& Toth, 
G. 2003, \apj, 591, L61. 
\bibitem[Parker(1958)]{parker} Parker, E. N. 1958m \apj, 123, 644.
\bibitem[Pauls et al.(1995)]{zankpauls} Pauls, H.L., Zank, G.P., \& 
Williams, L.L. 1995, \jgr, 100, 21595.
\bibitem[Pauls \& Zank(1996)]{pauls} Pauls, H.L., \& Zank, G.P. 1996, 
\jgr, 101, 17081.
\bibitem[Pogorolev \& Matsuda(2000)]{pogorolev} Pogorolev, N.V., \& 
Matsuda, T. 2000, \aap, 354, 697.
\bibitem[Powell et al.(1999)]{powell} Powell, K.G., Roe, P.L., Linde, T.J., 
Gombosi, T.I., and DeZeeuw, D.L. 1999, J.Comput.Phys., 154, 284.
\bibitem[Ratkiewicz et al.(1998)]{ratkiweicz98} Ratkiewicz, R., 
Barnes, A., Molvik, G. A., Spreiter, J. R., Stahara, S. S., Vinokur, 
M., Venkateswaran, S. 1998, Astron. \& Astrophysics, 335, 363.
\bibitem[Ratkiewicz, Barnes, \& Spreiter(2000)]{ratkiweicz00} 
Ratkiewicz, R., Barnes, A., \& Spreiter, J. R. 2000, \jgr, 105, 
25021. 
\bibitem[Ratkiewicz \& McKenzie(2003)]{ratkiweicz03} Ratkiewicz, R., 
\& McKenzie, J. F. 2003, \jgr, 108, 6.
\bibitem[Ratkiewicz \& Ben-Jaffel(2002)]{ratkiweiczjaffel} 
Ratkiewicz, R., \& Ben-Jaffel, L. 2002, \jgr, 107, 2.
\bibitem[e.g., Sahayanathan et al.(2003)] {jets} Sahayanathan, S., 
Misra, R. \& Kembhavi, A. K. 2003, \apj, 588, 77.
\bibitem[Smith(2001)]{smith} Smith, E. J. 2001, \jgr, 106, 
15819.
\bibitem[Suess(1990)]{suess} Suess, S.T. 1990, Rev.Geophys, 28, 97.
\bibitem[Washimi \& Tanaka(1996)]{washimi} Washimi, H., \& Tanaka, T. 
1996, Space Sci. Rev. 78, 85.
\bibitem[Washimi \& Tanaka(2001)]{washimiII} Washimi, H., \& Tanaka, 
T. 2001, {\it Adv. Space Res.}, 27, 509.
\bibitem[Winterhalter et al.(1994)]{daniel} Winterhalter, D., E.J. 
Smith, E. J., Burton, M. E., Murphy, N., \& McComas, D. J. 1994, 
\jgr, 99, 6667.
\bibitem[Velli(1994)]{velli} Velli, M. 1994, \apj, 432, L55. 
\bibitem[Zank et al.(1996)]{zank} Zank, G.P., Pauls, H.L., Williams, 
L.L., \& Hall, D. T. 1996, \jgr, 101, 21639. 
\end{thebibliography}
\end{document}